\documentclass[journal]{IEEEtran}

\usepackage{amsmath,amsfonts,amssymb}
\usepackage{tikz}
\usepackage{pgfplots}
\pgfplotsset{compat=1.18}
\usepackage{array}
\usepackage{multirow}
\usepackage[caption=false,font=normalsize,labelfont=sf,textfont=sf]{subfig}
\usepackage{stfloats}
\usepackage{url}
\usepackage{verbatim}
\usepackage{graphicx}
\usepackage{textcomp}
\usepackage[noend]{algpseudocode}
\usepackage{algorithm}
\usepackage{booktabs}
\usepackage{cite}
\usepackage[colorlinks=true,linkcolor=blue,citecolor=blue,urlcolor=blue]{hyperref}
\allowdisplaybreaks[2]

\begin{document}
\pagestyle{plain}

\title{RPolyPlan: A Unified Global Floorplanning and Legalization Framework for Rectilinear Soft Modules}

\author{Fuxing Huang, Wenxing Zhu, Xinning Liu, and Ziran Zhu
\thanks{Preliminary version of this paper was presented at the 2026 63rd ACM/IEEE Design Automation Conference (DAC), Long Beach, CA, USA, July 2026 \cite{DAC26}.}
\thanks{This work was partially supported by the National Natural Science Foundation of China under Grants 62674046 and 92464202.
(Corresponding author: Ziran Zhu).}
\thanks{Fuxing Huang, Xinning Liu, and Ziran Zhu are with the School of Integrated Circuits, Southeast University, Nanjing, China.}
\thanks{Wenxing Zhu is with the School of Mathematics and Statistics, Fuzhou University, Fuzhou, China.}
}

\maketitle
\thispagestyle{plain}

\begin{abstract}
Floorplanning with rectilinear soft modules improves area utilization and wirelength by allowing modules to adapt to rectilinear polygon (RPoly) shapes. Existing methods typically construct rectilinear polygons through incremental refinement after obtaining a rectangular global solution, which confines polygon exploration to the neighborhood induced by the initial floorplan. This paper presents RPolyPlan, a unified framework for outline-free and fixed-outline floorplanning that directly incorporates continuously deformable rectilinear polygons into global floorplanning. To enable direct polygon optimization, we introduce a structured continuous representation and a differentiable polygon shaping model that dynamically optimizes rectilinear module shapes based on overlaps, making continuous shape optimization feasible for gradient-based global floorplanning. An augmented Lagrangian formulation robustly handles the coupled inequality constraints, provides theoretical convergence guarantees, and reduces dependence on manual penalty-parameter tuning. During legalization, minimum-cost flow and rectangular decomposition determine the target areas of subrectangles, and horizontal and vertical constraint graphs encode their geometric relations. The resulting convex legalization formulation satisfies the area, shape, separation, connectivity, and outline constraints while minimizing either the chip area or the displacement from the global floorplanning solution. Experiments on the GSRC, MCNC, and ICCAD 2023 contest benchmarks demonstrate that RPolyPlan reduces HPWL over the evaluated state-of-the-art methods by 7\% and 6\% in fixed-outline and outline-free experiments, respectively. 
\end{abstract}


\section{Introduction}\label{sec:introduction}

With the continuous advancement of semiconductor technology, modern systems on chips (SoCs) integrate increasingly heterogeneous processing cores, memories, accelerators, and reusable intellectual property blocks. Hierarchical physical design is essential for managing this complexity, and floorplanning establishes the spatial organization on which subsequent placement and routing are performed~\cite{2003Parquet,tseng2024challenges}. By determining module locations, chip area, and the available routing space, a floorplan directly affects interconnect length, timing, power consumption, area utilization, and the convergence of downstream implementation. Consequently, obtaining a compact and legal floorplan with short wirelength is a fundamental prerequisite for achieving high physical-design quality.

Floorplanning is commonly considered under outline-free and fixed-outline settings. Outline-free floorplanning determines module locations and minimizes chip area while maintaining wirelength quality~\cite{kahng2000classical}. This formulation is suitable during early physical design when the chip area is not predetermined. Fixed-outline floorplanning instead places all modules within prescribed chip boundaries while satisfying geometric constraints and minimizing wirelength~\cite{2003Parquet}. It reflects modern hierarchical design, where the top-level outline is predetermined by architecture, packaging, and integration requirements. In both settings, restricting every module to a rectangle can leave fragmented regions unused.

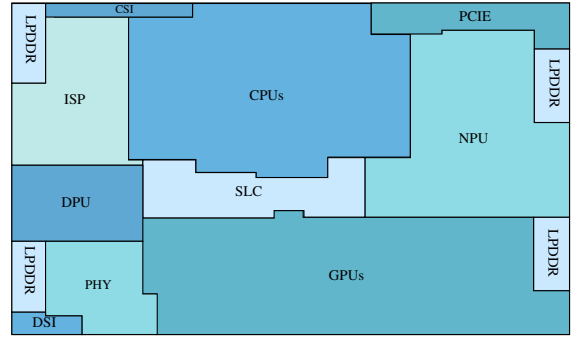
\begin{figure}[t]
\centering
\color{black}
\tikzset{every picture/.style={line width=0.75pt}} %

\resizebox{0.85\hsize}{!}{

\begin{tikzpicture}[x=0.75pt,y=0.75pt,yscale=-1,xscale=1]

\draw   (0,1.65) -- (504.83,1.65) -- (504.83,301.78) -- (0,301.78) -- cycle ;
\draw  [fill={rgb, 255:red, 190; green, 233; blue, 232 }  ,fill opacity=1 ] (105.66,143.74) -- (118.58,143.74) -- (118.58,148.54) -- (0,148.54) -- (0,74.12) -- (30.51,74.12) -- (30.51,13.5) -- (38.65,13.5) -- (38.65,14.33) -- (105.66,14.33) -- (105.66,143.74) -- cycle ;
\draw  [fill={rgb, 255:red, 99; green, 178; blue, 224 }  ,fill opacity=1 ] (105.66,143.33) -- (105.66,14.33) -- (163.57,14.33) -- (163.57,2.1) -- (166.44,2.1) -- (166.44,2.1) -- (325.35,2.1) -- (325.35,30.16) -- (360.77,30.16) -- (360.77,141.18) -- (286.1,141.18) -- (286.1,159.65) -- (221.01,159.65) -- (221.01,155.03) -- (166.44,155.03) -- (166.44,143.33) -- (105.66,143.33) -- cycle ;
\draw  [fill={rgb, 255:red, 142; green, 219; blue, 229 }  ,fill opacity=1 ] (473,26.26) -- (473,109.83) -- (504.83,109.83) -- (504.83,195.66) -- (501.96,195.66) -- (501.96,195.66) -- (319.6,195.66) -- (319.6,141.64) -- (360.77,141.64) -- (360.77,29.26) -- (389.24,29.26) -- (389.24,26.26) -- (473,26.26) -- cycle ;
\draw  [fill={rgb, 255:red, 95; green, 168; blue, 211 }  ,fill opacity=1 ] (163.57,2.1) -- (163.57,14.4) -- (30.51,14.4) -- (30.51,2.1) -- (163.57,2.1) -- cycle ;
\draw  [fill={rgb, 255:red, 98; green, 182; blue, 203 }  ,fill opacity=1 ] (389.24,29.71) -- (325.35,29.71) -- (325.35,1.65) -- (504.83,1.65) -- (504.83,43.36) -- (473,43.36) -- (473,26.26) -- (389.24,26.26) -- (389.24,29.71) -- cycle ;
\draw  [fill={rgb, 255:red, 202; green, 233; blue, 255 }  ,fill opacity=1 ] (473,43.36) -- (504.83,43.36) -- (504.83,109.83) -- (473,109.83) -- cycle ;
\draw  [fill={rgb, 255:red, 202; green, 233; blue, 255 }  ,fill opacity=1 ] (166.44,143.74) -- (166.44,155.15) -- (221.01,155.15) -- (221.01,159.65) -- (285.98,159.65) -- (285.98,141.64) -- (319.6,141.64) -- (319.6,195.66) -- (264.44,195.66) -- (264.44,189.62) -- (238.12,189.62) -- (238.12,196.37) -- (118.58,196.37) -- (118.58,143.74) -- (166.44,143.74) -- cycle ;
\draw  [fill={rgb, 255:red, 202; green, 233; blue, 255 }  ,fill opacity=1 ] (472.29,195.66) -- (504.83,195.66) -- (504.83,262.13) -- (472.29,262.13) -- cycle ;
\draw  [fill={rgb, 255:red, 98; green, 182; blue, 203 }  ,fill opacity=1 ] (263.72,189.62) -- (263.72,195.66) -- (472.29,195.66) -- (472.29,262.13) -- (504.83,262.13) -- (504.83,301.78) -- (130.9,301.78) -- (130.9,265.02) -- (117.86,265.02) -- (117.86,196.37) -- (237.4,196.37) -- (237.4,189.62) -- (263.72,189.62) -- cycle ;
\draw  [fill={rgb, 255:red, 95; green, 168; blue, 211 }  ,fill opacity=1 ] (0,148.54) -- (118.58,148.54) -- (118.58,217.38) -- (0,217.38) -- cycle ;
\draw  [fill={rgb, 255:red, 202; green, 233; blue, 255 }  ,fill opacity=1 ] (0,2.1) -- (30.51,2.1) -- (30.51,74.12) -- (0,74.12) -- cycle ;
\draw  [fill={rgb, 255:red, 202; green, 233; blue, 255 }  ,fill opacity=1 ] (0,217.38) -- (30.51,217.38) -- (30.51,281.56) -- (0,281.56) -- cycle ;
\draw  [fill={rgb, 255:red, 142; green, 219; blue, 229 }  ,fill opacity=1 ] (118.58,217.38) -- (118.58,265.02) -- (131.62,265.02) -- (131.62,301.78) -- (63.42,301.78) -- (63.42,284.9) -- (30.51,284.9) -- (30.51,217.38) -- (118.58,217.38) -- cycle ;
\draw  [fill={rgb, 255:red, 99; green, 178; blue, 224 }  ,fill opacity=1 ] (63.42,284.9) -- (63.42,301.78) -- (0,301.78) -- (0,281.56) -- (30.51,281.56) -- (30.51,284.9) -- (63.42,284.9) -- cycle ;

\draw (213.7,78.9) node [anchor=north west][inner sep=0.75pt]   [align=left] {CPUs};
\draw (403.27,117.66) node [anchor=north west][inner sep=0.75pt]   [align=left] {NPU};
\draw (200.75,165.68) node [anchor=north west][inner sep=0.75pt]   [align=left] {SLC};
\draw (285.59,241.45) node [anchor=north west][inner sep=0.75pt]   [align=left] {GPUs};
\draw (45.9,82.4) node [anchor=north west][inner sep=0.75pt]   [align=left] {ISP};
\draw (497.43,50.9) node [anchor=north west][inner sep=0.75pt]  [rotate=-90] [align=left] {LPDDR};
\draw (496.23,204.7) node [anchor=north west][inner sep=0.75pt]  [rotate=-90] [align=left] {LPDDR};
\draw (23.39,11.14) node [anchor=north west][inner sep=0.75pt]  [rotate=-90] [align=left] {LPDDR};
\draw (22.39,220.45) node [anchor=north west][inner sep=0.75pt]  [rotate=-90] [align=left] {LPDDR};
\draw (62.57,251.25) node [anchor=north west][inner sep=0.75pt]  [font=\small] [align=left] {\begin{minipage}[lt]{21.6pt}\setlength\topsep{0pt}
\begin{center}
PHY
\end{center}

\end{minipage}};
\draw (17.26,284.56) node [anchor=north west][inner sep=0.75pt]   [align=left] {DSI};
\draw (92.18,2) node [anchor=north west][inner sep=0.75pt]  [font=\footnotesize] [align=left] {CSI};
\draw (403.56,7.38) node [anchor=north west][inner sep=0.75pt]   [align=left] {PCIE};
\draw (43.4,176.01) node [anchor=north west][inner sep=0.75pt]   [align=left] {DPU};

\end{tikzpicture}
}
 \caption{A chip floorplan diagram shows that the functional modules in a modern SoC exhibit complex rectilinear polygon shapes.}
  \label{fig:zen4core}
\end{figure}

Rectilinear soft modules permit flexible orthogonal boundaries, improving irregular-space utilization and shortening interconnects~\cite{2024NTUFP,cadcontest23}. As illustrated by the chip floorplan in Figure~\ref{fig:zen4core}, such boundaries arise naturally in complex SoC designs. Fully exploiting this flexibility requires polygonal degrees of freedom to participate in global floorplanning, where module locations and interconnect cost are jointly optimized. Existing approaches instead obtain an initial rectangular floorplan and subsequently form rectilinear modules through incremental refinement, causing the reachable polygons to remain constrained by the initial solution. To our knowledge, directly optimizing rectilinear polygons during global floorplanning has not been addressed. This unexplored coupling constitutes the central challenge considered in this work.

\subsection{Previous Work}

Table~\ref{tab:capability-comparison} summarizes the capabilities supported by previous floorplanning methods. Existing methods can be broadly classified into metaheuristic, learning-based, and analytical approaches. Metaheuristic methods commonly employ simulated annealing~\cite{kirkpatrick1983SA} to perturb topological representations, including slicing trees, sequence pairs, O-trees, B*-trees, corner block lists, and TCG-S~\cite{1986Stree,1995SP,1999OTree,2000BTree,2000CBL,2002TCGS}. Each candidate topology is packed and evaluated, enabling flexible exploration without requiring a differentiable objective. Fixed-outline extensions further incorporate outline penalties, deferred shape decisions, or multilevel refinement to improve feasibility for large instances~\cite{2003Parquet,2008IMF,2010DeFer}. Nevertheless, the number of perturbations required for high-quality solutions grows rapidly with problem size, and these representations were primarily developed for rectangular modules. Earlier rectilinear extensions either prescribe a limited set of L- or T-shaped modules or decompose known rectilinear shapes into abutting rectangles~\cite{kang1997general,xu1998rectilinear,wu2003rectilinear}. Their global search therefore cannot freely coordinate module placement with deformable polygon geometry. TOFU~\cite{2023tofu} refines a near-legal rectangular floorplan by greedily redistributing whitespace and incrementally forming rectilinear modules. Because this refinement begins from an existing solution, its achievable shapes and inter-module relations remain constrained by the initial rectangular floorplan.

\begingroup
\hypersetup{citecolor=black}
\begin{table}[t]
  \centering
  \caption{Capability comparison of representative floorplanning methods.}
  \label{tab:capability-comparison}
  \resizebox{\columnwidth}{!}{
    \begin{tabular}{@{}lcccc@{}}
      \toprule
      Works
       & \shortstack{Outline                                    \\ Free}
       & \shortstack{Fixed                                                       \\Outline}
       & \shortstack{Local Rectilinear                                           \\Optimization}
       & \shortstack{Global Rectilinear                                           \\Optimization} \\
      \midrule
      \cite{
        1986Stree,
        1995SP,
        2000BTree,
        2000CBL,
        2002TCGS,
        2020rlsa,
        2022goodfp
      }
       & $\checkmark$               & $\times$     & $\times$     & $\times$     \\
      \cite{
        2006Analytical,
        2014F-FM,
        2022kdd,
        2022PeF,
        2023rotationfp,
        2023SDP
      }
       & $\triangle$                   & $\checkmark$ & $\times $    & $\times$     \\
      \cite{
        2003Parquet,
        2008IMF,
        2010DeFer,
        2011UFO
      }
       & $\checkmark$               & $\checkmark$ & $\times$     & $\times$     \\
      \cite{
        xu1998rectilinear,
        wu2003rectilinear,
        2023tofu
      }
       & $\checkmark$               & $\times$     & $\checkmark$ & $\times$     \\
      \cite{kahng2000classical, 2024NTUFP}
       & $\triangle$                   & $\checkmark$ & $\checkmark$ & $\times$     \\
      \cite{
        lp, 2024jigsawplanner
      }
       & $\checkmark$               & $\checkmark$ & $\checkmark$ & $\times$     \\
      Ours
       & $\checkmark$               & $\checkmark$ & $\checkmark$ & $\checkmark$ \\
      \bottomrule
    \end{tabular}
  }
  \vspace{2pt}

  \parbox{0.98\hsize}{\scriptsize
    Local and Global denote incremental rectilinear polygon optimization based on an existing floorplan and global rectilinear polygon optimization, respectively. $\checkmark$, $\triangle$, and $\times$ indicate full, partial, and unsupported capabilities.}
\end{table}
\endgroup

Learning-based methods use learned policies to guide floorplanning search. Reinforcement learning has been applied to local-search operations for fixed-outline floorplanning~\cite{2020rlsa}. GoodFloorplan~\cite{2022goodfp} combines graph convolutional networks with reinforcement learning to guide sequential construction, while a generalizable floorplanner~\cite{2022kdd} integrates a corner block list with hypergraph embedding. Although learned guidance can reduce manually designed search rules, its performance depends on the training distribution, reward design, and inference budget. Moreover, these methods learn discrete perturbations, module orders, or rectangular packing decisions rather than continuously deformable polygons. They therefore do not jointly optimize rectilinear shapes, module locations, and wirelength in the global stage, leaving subsequent polygon refinement dependent on the initial floorplan.

Analytical methods formulate floorplanning as continuous optimization and are generally more scalable than repeated packing-based search~\cite{2006Analytical,2014F-FM,2022PeF,2023rotationfp,2023SDP}. They typically separate a smooth global floorplanning stage from legalization, but their global variables describe rectangular module locations and dimensions. Although convex and force-directed formulations improve scalability, their relaxed rectangular geometry can produce overlap or outline violations that must be repaired afterward. Recent methods introduce rectilinear flexibility only in downstream refinement. JigsawPlanner~\cite{2024jigsawplanner} employs a PeF-based global stage~\cite{2022PeF} with square area abstractions, then partitions the layout and modules into grids and applies hierarchical assignment and cellular-automata refinement to construct complex shapes. Thus, its global stage does not optimize polygon geometry, while the subsequent discretization trades geometric fidelity for runtime. ICCAD'24~\cite{2024NTUFP} optimizes shape-adjustable rectangles during global floorplanning using ReLU-inspired derivative handling, then migrates area through an overlap graph and incrementally refines rectangular bounding boxes into legal rectilinear contours. Therefore, both methods determine polygonal shapes only after obtaining a nonpolygonal global solution. This separation constrains polygon exploration to the neighborhood induced by that solution and prevents global floorplanning from jointly exploiting module placement, wirelength, and rectilinear shape freedom.

\subsection{Our Contributions}

To address this gap, we develop a unified framework that directly couples continuous rectilinear shape optimization with convex legalization. Rather than introducing polygonal flexibility only after a rectangular global solution has been obtained, the proposed method represents rectilinear shapes with continuous variables during global floorplanning and subsequently produces a legal floorplan through area redistribution and convex optimization. Our main contributions are summarized as follows:
\begin{itemize}
  \item We propose a novel representation that formalizes complex rectilinear shapes using a structured set of continuous variables, together with a differentiable polygon shaping model that dynamically optimizes module shapes based on overlaps. This representation makes continuous rectilinear shape optimization during global floorplanning feasible for gradient-based methods.
  \item We develop an augmented Lagrangian formulation that robustly and efficiently handles multiple coupled inequality constraints, particularly those induced by the differentiable polygon shaping model. This formulation provides theoretical convergence guarantees and reduces the dependence on manual hyperparameter tuning inherent in traditional penalty methods.
  \item We propose a convex-based legalization flow that combines minimum-cost flow, rectangular decomposition, and constraint graphs to determine subrectangle areas and geometric relations. The resulting convex formulation satisfies the area, shape, separation, connectivity, and outline constraints while minimizing the chip area or displacement from the global floorplanning solution.
  \item Experiments on GSRC~\cite{GSRCbench}, MCNC~\cite{MCNCbench}, and ICCAD 2023 contest benchmarks~\cite{cadcontest23} validate both optimization modes. Compared with the evaluated SOTA methods, our method reduces HPWL by 7\% and 6\% in fixed-outline and outline-free optimization, respectively, while minimizing chip area and achieving low whitespace under the outline-free setting. The runtime remains practical across all benchmarks.
\end{itemize}

\section{Preliminaries}\label{sec:pre}
This section introduces the rectilinear floorplanning problem, including the module types, geometric constraints, and the fixed-outline and outline-free optimization modes. The exact HPWL model and its smooth approximation used in global floorplanning are then presented.
\subsection{Problem Formulation}

The floorplanning problem with rectilinear soft modules can be formulated as a joint optimization of hypergraph placement and geometric shaping.
Given a hypergraph $\mathcal{H} = (\mathcal{V}, \mathcal{E}, \mathcal{F})$, the vertex set $\mathcal{V}=\{v_1,v_2,\dots,v_m\}$ represents the circuit modules, the hyperedge set $\mathcal{E}=\{e_1,e_2,\dots,e_n\}$ represents the nets interconnecting these modules, and the placement region $\mathcal{F}$ is a rectangle with width $W$ and height $H$.

{
The module set consists of soft modules $\mathcal V_s$, hard modules $\mathcal V_h$, and preplaced modules $\mathcal V_p$, with $\mathcal V=\mathcal V_s\mathbin{\dot\cup}\mathcal V_h\mathbin{\dot\cup}\mathcal V_p$. Soft modules have movable positions and flexible rectilinear shapes, hard modules have movable positions and fixed dimensions, and preplaced modules have fixed positions and dimensions.
}

A legal placement must satisfy two primary constraints:
(i) the \textbf{outline constraint}, which requires all modules to be entirely contained within $\mathcal{F}$, and
(ii) the \textbf{non-overlapping constraint}, which prohibits any two modules from overlapping.

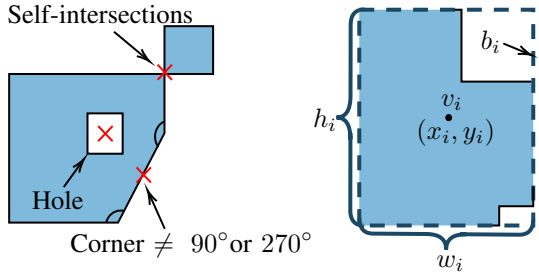
\begin{figure}[htbp]
\centering

\tikzset{every picture/.style={line width=0.75pt}}

\begin{tikzpicture}[x=0.75pt,y=0.75pt,yscale=-1.1,xscale=1]

\draw  [color={rgb, 255:red, 27; green, 73; blue, 101 }  ,draw opacity=1 ][line width=1.5]  (178.4,8.3) .. controls (173.73,8.29) and (171.4,10.62) .. (171.39,15.29) -- (171.32,47.59) .. controls (171.31,54.26) and (168.97,57.58) .. (164.3,57.57) .. controls (168.97,57.58) and (171.29,60.92) .. (171.28,67.59)(171.28,64.59) -- (171.21,99.89) .. controls (171.2,104.56) and (173.52,106.89) .. (178.19,106.9) ;
\draw  [draw opacity=0][fill={rgb, 255:red, 95; green, 168; blue, 211 }  ,fill opacity=0.8 ][line width=0.75]  (175.71,8.4) -- (227.01,8.4) -- (227.01,41.2) -- (263.03,41.2) -- (263.03,98.06) -- (245.94,98.06) -- (245.94,106.85) -- (175.71,106.85) -- cycle ;
\draw  [color={rgb, 255:red, 0; green, 0; blue, 0 }  ,draw opacity=1 ][fill={rgb, 255:red, 0; green, 0; blue, 0 }  ,fill opacity=1 ] (219.37,57.63) .. controls (219.37,56.98) and (219.92,56.45) .. (220.59,56.45) .. controls (221.27,56.45) and (221.81,56.98) .. (221.81,57.63) .. controls (221.81,58.27) and (221.27,58.8) .. (220.59,58.8) .. controls (219.92,58.8) and (219.37,58.27) .. (219.37,57.63) -- cycle ;
\draw  [color={rgb, 255:red, 27; green, 73; blue, 101 }  ,draw opacity=1 ][dash pattern={on 5.63pt off 4.5pt}][line width=1.5]  (175.71,8.4) -- (263.03,8.4) -- (263.03,106.85) -- (175.71,106.85) -- cycle ;
\draw    (248,23.8) -- (260.21,28.39) ;
\draw [shift={(262.09,29.09)}, rotate = 200.58] [color={rgb, 255:red, 0; green, 0; blue, 0 }  ][line width=0.75]    (6.56,-1.97) .. controls (4.17,-0.84) and (1.99,-0.18) .. (0,0) .. controls (1.99,0.18) and (4.17,0.84) .. (6.56,1.97)   ;
\draw    (78.09,108.36) -- (70.3,92.49) ;
\draw [shift={(69.41,90.7)}, rotate = 63.83] [color={rgb, 255:red, 0; green, 0; blue, 0 }  ][line width=0.75]    (6.56,-1.97) .. controls (4.17,-0.84) and (1.99,-0.18) .. (0,0) .. controls (1.99,0.18) and (4.17,0.84) .. (6.56,1.97)   ;
\draw    (52.62,18.44) -- (68.9,30.8) ;
\draw [shift={(70.5,32.01)}, rotate = 217.19] [color={rgb, 255:red, 0; green, 0; blue, 0 }  ][line width=0.75]    (6.56,-1.97) .. controls (4.17,-0.84) and (1.99,-0.18) .. (0,0) .. controls (1.99,0.18) and (4.17,0.84) .. (6.56,1.97)   ;
\draw  [fill={rgb, 255:red, 95; green, 168; blue, 211 }  ,fill opacity=0.8 ] (77.95,15.91) -- (102.36,15.91) -- (102.36,37.72) -- (77.95,37.72) -- cycle ;
\draw  [fill={rgb, 255:red, 95; green, 168; blue, 211 }  ,fill opacity=0.8 ] (77.95,37.72) -- (77.95,64.73) -- (54.64,105.53) -- (0,105.53) -- (0,37.72) -- cycle ;
\draw  [color={rgb, 255:red, 255; green, 0; blue, 0 }  ,draw opacity=1 ] (74.49,33.48) -- (81.71,40.39)(81.72,33.47) -- (74.49,40.4) ;
\draw  [fill={rgb, 255:red, 255; green, 255; blue, 255 }  ,fill opacity=1 ] (39.32,55.69) -- (56.96,55.69) -- (56.96,74.1) -- (39.32,74.1) -- cycle ;
\draw  [color={rgb, 255:red, 255; green, 0; blue, 0 }  ,draw opacity=1 ] (44.68,61.48) -- (51.9,68.4)(51.91,61.48) -- (44.68,68.4) ;
\draw  [color={rgb, 255:red, 255; green, 0; blue, 0 }  ,draw opacity=1 ] (63.93,80.33) -- (71.15,87.24)(71.16,80.32) -- (63.93,87.25) ;
\draw    (27.41,89.31) -- (36.47,77.9) ;
\draw [shift={(37.71,76.33)}, rotate = 128.43] [color={rgb, 255:red, 0; green, 0; blue, 0 }  ][line width=0.75]    (6.56,-1.97) .. controls (4.17,-0.84) and (1.99,-0.18) .. (0,0) .. controls (1.99,0.18) and (4.17,0.84) .. (6.56,1.97)   ;
\draw  [color={rgb, 255:red, 27; green, 73; blue, 101 }  ,draw opacity=1 ][line width=1.5]  (175.8,103.5) .. controls (175.77,108.17) and (178.08,110.52) .. (182.75,110.55) -- (209.35,110.73) .. controls (216.02,110.78) and (219.33,113.13) .. (219.3,117.8) .. controls (219.33,113.13) and (222.68,110.82) .. (229.35,110.86)(226.35,110.84) -- (255.95,111.05) .. controls (260.62,111.08) and (262.97,108.76) .. (263,104.09) ;
\draw  [draw opacity=0] (48.75,105.3) .. controls (49,104.49) and (49.39,103.7) .. (49.94,102.99) .. controls (51.8,100.53) and (54.88,99.54) .. (57.69,100.23) -- (55.91,107.52) -- cycle ; \draw   (48.75,105.3) .. controls (49,104.49) and (49.39,103.7) .. (49.94,102.99) .. controls (51.8,100.53) and (54.88,99.54) .. (57.69,100.23) ;  
\draw  [draw opacity=0] (74.9,71.05) .. controls (74.36,70.39) and (73.93,69.63) .. (73.63,68.78) .. controls (72.34,65.08) and (74.13,61.05) .. (77.63,59.48) -- (80.71,66.32) -- cycle ; \draw   (74.9,71.05) .. controls (74.36,70.39) and (73.93,69.63) .. (73.63,68.78) .. controls (72.34,65.08) and (74.13,61.05) .. (77.63,59.48) ;  
\draw    (223.5,8.4) -- (227.01,8.4) -- (227.01,41.2) -- (262.03,41.2);
\draw    (263,89.15) -- (263.03,98.06) -- (245.94,98.06) -- (245.94,106.85) -- (242.5,106.85) ;

\draw (215.26,45.89) node [anchor=north west][inner sep=0.75pt]    {$v_{i}$};
\draw (203,57) node [anchor=north west][inner sep=0.75pt]  {$( x_{i} ,y_{i})$};
\draw (235.26,16.43) node [anchor=north west][inner sep=0.75pt]    {$b_{i}$};
\draw (151,52) node [anchor=north west][inner sep=0.75pt]    {$h_{i}$};
\draw (213,119) node [anchor=north west][inner sep=0.75pt]    {$w_{i}$};
\draw (9.99,89.88) node [anchor=north west][inner sep=0.75pt]  [font=\normalsize] [align=left] {Hole};
\draw (-2.4,5) node [anchor=north west][inner sep=0.75pt]  [font=\normalsize] [align=left] {Self-intersections};
\draw (29.32,108.53) node [anchor=north west][inner sep=0.75pt]  [font=\normalsize] [align=left] {Corner $\displaystyle \neq \ 90^{\circ }$or $\displaystyle 270^{\circ }$ };

\end{tikzpicture}
\caption{Shape constraints.}
\label{fig:shape}
\end{figure}

Each module $v_i$ is modeled with a minimum bounding box $b_i = (x_i, y_i, w_i, h_i)$, where $(x_i, y_i)$ denotes the center coordinates, and $w_i$ and $h_i$ represent the width and height, respectively.
The coordinate of $v_i$ is determined by $(x_i, y_i)$, which also serves as the connection pin for its associated nets.
In addition, as illustrated in Figure~\ref{fig:shape}, each soft module $v_i\in\mathcal V_s$ must satisfy the following shape constraints~\cite{2024NTUFP}:
\begin{itemize}
    \item \textbf{Simplicity:} $v_i$ must be a simple rectilinear polygon with axis-aligned edges, interior angles of 90° or 270°, no holes, and no self-intersections.
    \item \textbf{Minimum area:} The area of $v_i$ must be no smaller than its specified minimum value $A_i^{\min}$.
    \item \textbf{Aspect ratio:} The aspect ratio of $v_i$, defined as the ratio between the height and width of $b_i$, must lie within the range $[\text{AR}_i^l, \text{AR}_i^u]$.
    \item \textbf{Rectangle ratio:} The ratio between the area of $v_i$ and the area of $b_i$ must fall within $[\text{RR}_i^{\min}, 1]$.
    \item \textbf{Minimum inner width:} The inner width of $v_i$ must be no smaller than the specified value $\omega^{\min}$.
\end{itemize}

The optimization objective is to determine the coordinates and shapes of all movable modules such that the aforementioned constraints are satisfied. We consider two optimization modes. In the fixed-outline mode, $W$ and $H$ are prescribed constants, and the objective is to minimize the total HPWL within $\mathcal F$. In the outline-free mode, $W$ and $H$ are optimization variables, and chip area is minimized together with HPWL. Accordingly, the floorplanning problem with rectilinear soft modules can be mathematically formulated as
\begin{subequations}\label{eq:obj1}
    \begin{align}
        \min \quad & 
        \displaystyle\sum_{e\in\mathcal E}\mathcal W_e(\mathbf x,\mathbf y)+\eta WH
        \label{eq:obj1:obj}\\
        \text{s.t.} \quad
                   & v_i \cap v_j = \varnothing, \quad \forall v_i,v_j\in\mathcal V, i \neq j,\label{eq:obj1:overlap}\\
                   & v_i \subseteq \mathcal{F}, \quad \forall v_i\in\mathcal V, \label{eq:obj1:outline}\\
                   & v_i \;\text{is simple}, \quad \forall v_i\in\mathcal V_s, \label{eq:obj1:rectilinear}\\
                   & \text{Area}(v_i) \geq A_i^{\min}, \quad \forall v_i\in\mathcal V_s, \label{eq:obj1:area}\\
                   & \text{AR}_i^l \leq \tfrac{h_i}{w_i} \leq \text{AR}_i^u, \quad \forall v_i\in\mathcal V_s, \label{eq:obj1:aspect}\\
                   & \text{RR}_i^{\min} \leq \tfrac{\text{Area}(v_i)}{w_i h_i} \leq 1.0, \quad \forall v_i\in\mathcal V_s, \label{eq:obj1:rect-ratio}\\
                   & (w_i,h_i)=(\bar w_i,\bar h_i),
                   \quad \forall v_i\in\mathcal V_h, \label{eq:obj1:hard}\\
                   & b_i=\bar b_i, \quad \forall v_i\in\mathcal V_p, \label{eq:obj1:fixed}\\
                   & v_i \;\text{has inner width at least } \omega^{\min}, \quad \forall v_i\in\mathcal V_s. \label{eq:obj1:inner-width}
    \end{align}
\end{subequations}
The two modes share the same constraints. The weight $\eta$ balances HPWL and chip area. For fixed-outline floorplanning, $W$ and $H$ are prescribed and we set $\eta=0$. For outline-free floorplanning, $W$ and $H$ are optimization variables and $\eta>0$. The symbols $\bar w_i$, $\bar h_i$, and $\bar b_i$ denote the given hard and preplaced module geometry.
Constraints~\eqref{eq:obj1:overlap}--\eqref{eq:obj1:outline} ensure placement legality, constraints~\eqref{eq:obj1:rectilinear}--\eqref{eq:obj1:rect-ratio} guarantee the geometric feasibility of soft module shapes, and constraints~\eqref{eq:obj1:hard}--\eqref{eq:obj1:fixed} preserve hard and preplaced module geometry.

\subsection{Wirelength Model}\label{subsec:wirelength-model}
{
Half-perimeter wirelength (HPWL) is widely used to estimate wirelength in floorplanning. Let $\mathbf x=\{x_i\mid v_i\in\mathcal V\}$ and $\mathbf y=\{y_i\mid v_i\in\mathcal V\}$ denote the module coordinates. For a net $e\in\mathcal E$, its HPWL is defined as
\begin{equation}
\begin{aligned}
\mathcal W_e(\mathbf x,\mathbf y):={}&\!
\max_{v_i\in e}x_i-\min_{v_i\in e}x_i\!+\!\max_{v_i\in e}y_i-\min_{v_i\in e}y_i,
\end{aligned}
\label{eq:hpwl-model}
\end{equation}
The two terms in Eq.~\eqref{eq:hpwl-model} measure the horizontal and vertical spans of $e$ using module centers as connection pins. Since HPWL is non-differentiable, we adopt the weighted-average (WA) wirelength model~\cite{hsu2013tsv} with coefficient $\gamma$ for gradient-based global floorplanning. Its horizontal component is
\begin{equation}
\hat{\mathcal W}_e^x(\mathbf x)=
\frac{\sum_{v_i\in e}x_i\exp(x_i/\gamma)}
{\sum_{v_i\in e}\exp(x_i/\gamma)}
-\frac{\sum_{v_i\in e}x_i\exp(-x_i/\gamma)}
{\sum_{v_i\in e}\exp(-x_i/\gamma)},
\label{eq:wa-wirelength-x}
\end{equation}
and $\hat{\mathcal W}_e^y(\mathbf y)$ is defined similarly. The smooth wirelength of net $e$ is
\begin{equation}
\hat{\mathcal W}_e(\mathbf x,\mathbf y)
=\hat{\mathcal W}_e^x(\mathbf x)+\hat{\mathcal W}_e^y(\mathbf y).
\label{eq:wa-wirelength-model}
\end{equation}
A smaller $\gamma>0$ provides a more accurate approximation of HPWL. Global floorplanning uses the WA model, while the final solution is evaluated using exact HPWL.
}

\section{Proposed Method}\label{sec:app}
\input{include/Figure/proposed_method_flow}
The overall framework is illustrated in Figure~\ref{fig:proposed-method-flow}. Given the module and netlist information, continuous optimization performs position movement, polygon shaping, bounding-box area adjustment, and aspect-ratio adjustment. Polygon shape determination then converts the optimized solution into explicit rectilinear polygons. During legalization, minimum-cost flow plans area redistribution between neighboring polygons, and rectangular decomposition generates subrectangles with their target areas. Constraint graph construction encodes the horizontal and vertical geometric relations, after which convex legalization adjusts the rectangle locations and dimensions while minimizing chip area or displacement. Finally, the optimized subrectangles are merged to produce the legalized rectilinear floorplan.

\subsection{Global Floorplanning}\label{sec:globalFloorplan}
Global floorplanning formulates the floorplanning task as a continuous optimization problem, in which both the coordinates and shapes of modules are treated as decision variables.
Recent advances in \textit{rectilinear soft modules} have introduced flexible shape adaptation, allowing modules to deform to alleviate overlap with neighboring modules and potentially achieve shorter wirelength.
However, representing such flexible geometries with a compact and differentiable set of continuous variables remains challenging.
Although conventional rectangular modules can be described by only a few variables (e.g., coordinates, width, and height), rectilinear modules possess significantly higher degrees of freedom, rendering direct optimization computationally intractable.
As a result, previous studies~\cite{2024NTUFP,2024jigsawplanner} have typically approximated rectilinear modules as rectangles to maintain computational efficiency.

\begin{figure}[htbp]

\centering
\tikzset{every picture/.style={line width=0.75pt}}

\begin{tikzpicture}[x=0.75pt,y=0.75pt,yscale=-1,xscale=1]

\draw  [draw opacity=0][fill={rgb, 255:red, 95; green, 168; blue, 211 }  ,fill opacity=0.8 ][line width=0.75]  (0,45.2) -- (42,45.2) -- (42,85) -- (71.49,85) -- (71.49,104.25) -- (57.5,104.25) -- (57.5,129.25) -- (0,129.25) -- cycle ;
\draw  [draw opacity=0][fill={rgb, 255:red, 190; green, 233; blue, 232 }  ,fill opacity=0.8 ][line width=0.75]  (42,45.2) -- (42,25) -- (112,25) -- (112,85) -- (71.49,85) -- (71.49,45.2) -- cycle ;
\draw  [draw opacity=0][fill={rgb, 255:red, 202; green, 233; blue, 255 }  ,fill opacity=0.8 ][line width=0.75]  (71.49,104.25) -- (86.99,104.25) -- (86.99,144.25) -- (57.5,144.25) -- (57.5,129.25) -- (71.49,129.25) -- cycle ;
\draw  [draw opacity=0][fill={rgb, 255:red, 162; green, 214; blue, 226 }  ,fill opacity=0.8 ][line width=0.75]  (42,45.2) -- (71.49,45.2) -- (71.49,85) -- (42,85) -- cycle ;
\draw  [draw opacity=0][fill={rgb, 255:red, 127; green, 188; blue, 224 }  ,fill opacity=0.8 ][line width=0.75]  (57.5,104.25) -- (71.49,104.25) -- (71.49,129.25) -- (57.5,129.25) -- cycle ;
\draw    (118.4,25.4) .. controls (85.7,42.34) and (73.95,21.25) .. (58.54,56.38) ;
\draw [shift={(57.83,58.02)}, rotate = 292.85] [color={rgb, 255:red, 0; green, 0; blue, 0 }  ][line width=0.75]    (6.56,-1.97) .. controls (4.17,-0.84) and (1.99,-0.18) .. (0,0) .. controls (1.99,0.18) and (4.17,0.84) .. (6.56,1.97)   ;
\draw    (118.4,25.4) .. controls (117.21,84.8) and (82.7,58.74) .. (64.55,111.38) ;
\draw [shift={(64,113)}, rotate = 288.18] [color={rgb, 255:red, 0; green, 0; blue, 0 }  ][line width=0.75]    (6.56,-1.97) .. controls (4.17,-0.84) and (1.99,-0.18) .. (0,0) .. controls (1.99,0.18) and (4.17,0.84) .. (6.56,1.97)   ;
\draw  [color={rgb, 255:red, 0; green, 0; blue, 0 }  ,draw opacity=1 ][fill={rgb, 255:red, 95; green, 168; blue, 211 }  ,fill opacity=0.8 ][line width=0.75]  (173.33,45.2) -- (215.33,45.2) -- (215.33,73.2) -- (244.83,73.2) -- (244.83,121.75) -- (230.83,121.75) -- (230.83,129.25) -- (173.33,129.25) -- cycle ;
\draw  [color={rgb, 255:red, 0; green, 0; blue, 0 }  ,draw opacity=1 ][fill={rgb, 255:red, 190; green, 233; blue, 232 }  ,fill opacity=0.8 ][line width=0.75]  (215.33,73.2) -- (215.33,25) -- (285.33,25) -- (285.33,85) -- (244.83,85) -- (244.83,73.2) -- cycle ;
\draw  [color={rgb, 255:red, 0; green, 0; blue, 0 }  ,draw opacity=1 ][fill={rgb, 255:red, 202; green, 233; blue, 255 }  ,fill opacity=0.8 ][line width=0.75]  (244.83,104.25) -- (260.33,104.25) -- (260.33,144.25) -- (230.83,144.25) -- (230.83,121.75) -- (244.83,121.75) -- cycle ;
\draw  [color={rgb, 255:red, 27; green, 73; blue, 101 }  ,draw opacity=1 ][dash pattern={on 5.63pt off 4.5pt}][line width=1.5]  (0,45.2) -- (71.49,45.2) -- (71.49,129.25) -- (0,129.25) -- cycle ;
\draw  [color={rgb, 255:red, 27; green, 73; blue, 101 }  ,draw opacity=1 ][dash pattern={on 5.63pt off 4.5pt}][line width=1.5]  (42,25) -- (112,25) -- (112,85) -- (42,85) -- cycle ;
\draw  [color={rgb, 255:red, 27; green, 73; blue, 101 }  ,draw opacity=1 ][dash pattern={on 5.63pt off 4.5pt}][line width=1.5]  (57.5,104.25) -- (86.99,104.25) -- (86.99,144.25) -- (57.5,144.25) -- cycle ;

\draw (48.4,59.89) node [anchor=north west][inner sep=0.75pt]  {$0.7$};
\draw (57.25,114.03) node [anchor=north west][inner sep=0.75pt]  [font=\footnotesize]  {$0.3$};
\draw (31.17,82.13) node [anchor=north west][inner sep=0.75pt]    {$i$};
\draw (75.33,43.79) node [anchor=north west][inner sep=0.75pt]    {$j$};
\draw (74.25,126.85) node [anchor=north west][inner sep=0.75pt]    {$k$};
\draw (204.5,82.13) node [anchor=north west][inner sep=0.75pt]    {$i$};
\draw (248.67,43.79) node [anchor=north west][inner sep=0.75pt]    {$j$};
\draw (247.58,126.85) node [anchor=north west][inner sep=0.75pt]    {$k$};
\draw (70,10) node [anchor=north west][inner sep=0.75pt]   [align=left] {Shaping Proportion};

\draw (45,150) node [anchor=north west][inner sep=0.75pt]   [align=left][font=\large]  {(a)};
\draw (218,150) node [anchor=north west][inner sep=0.75pt]   [align=left][font=\large]  {(b)};
\end{tikzpicture}
\caption{Illustration of polygon representation. (a) Three overlapping bounding boxes of modules and the shaping proportion of overlap areas; (b) The rectilinear modules corresponding to (a).}
\label{fig:polygonshaping}
\end{figure}
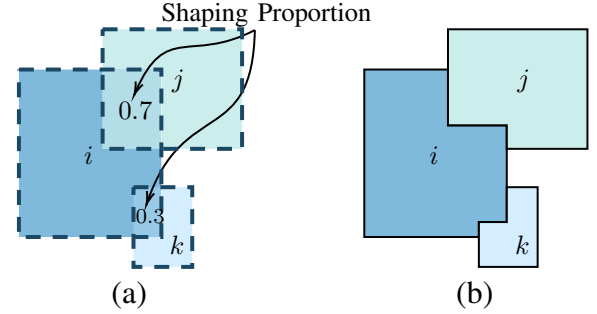

\subsubsection{Rectilinear Polygon Representation}

To enable differentiable optimization of polygonal shapes, we propose a formal representation that models each rectilinear polygon using a structured set of continuous variables.
As illustrated in Figure~\ref{fig:polygonshaping}, a polygon is characterized by its \textit{minimum bounding box} and a set of \textit{shaping proportions} that describe how overlapping areas are distributed among adjacent modules. Let $\mathcal V_m=\mathcal V_s\cup\mathcal V_h$ denote the set of movable modules.

\textbf{Definition 1}~(Bounding Box Variables).
{
\textit{For each movable module $v_i\in\mathcal V_m$, let $a_i$ denote its bounding box area and let $\mathrm{ar}_i=h_i/w_i$ denote its aspect ratio. The bounding box variables are defined as
\begin{equation}
    \mathbf v:=\{(x_i,y_i,a_i,\mathrm{ar}_i)\mid v_i\in\mathcal V_m\}
    \in\mathbb R^{4|\mathcal V_m|}.
    \label{eq:bounding-box-variables}
\end{equation}
The width and height are recovered by
\begin{equation}
    w_i=\sqrt{\frac{a_i}{\mathrm{ar}_i}},
    \qquad
    h_i=\sqrt{a_i\mathrm{ar}_i},
    \label{eq:area-ar-conversion}
\end{equation}
which guarantees $w_i h_i=a_i$ and $h_i/w_i=\mathrm{ar}_i$. Accordingly, $\mathbf v$ determines the geometric bounding box set
\begin{equation}
    \mathbf b(\mathbf v)
    :=\{b_i(\mathbf v)=(x_i,y_i,w_i,h_i)\mid v_i\in\mathcal V_m\}.
    \label{eq:recovered-bboxes}
\end{equation}
}
}
{
The bounding box variables $\mathbf v$ explicitly control the module location, area, and aspect ratio. The transformation in Eq.~\eqref{eq:area-ar-conversion} automatically satisfies $w_i h_i=a_i$ and $h_i/w_i=\mathrm{ar}_i$, while providing the width and height required by subsequent geometric models. This parameterization therefore exposes the area and aspect ratio to continuous optimization without introducing additional equality constraints between $w_i$ and $h_i$.
}

\textbf{Definition 2}~(Shaping Variables).
\textit{For each pair of movable modules $(v_i,v_j)\in\mathcal V_m\times\mathcal V_m$, we introduce a shaping variable $r_{ij}$ that determines the proportion of the overlap contributed by module $v_i$.
Let $\mathbf{r} := \{r_{ij} \mid v_i,v_j\in\mathcal V_m,\ i<j\} \in \mathbb{R}^{|\mathcal V_m|(|\mathcal V_m|-1)/2}$ denote the set of all shaping variables.
To ensure these proportions lie within the interval $[0, 1]$, we apply a nonlinear sigmoid~\cite{sigmoid} based shaping proportion function:
\begin{equation}
    \nonumber
    \sigma_{ij}(\mathbf{r}) :=
    \begin{cases}
     \dfrac{1}{1 + e^{-r_{ij}}}, & \text{if } i < j, \\[6pt]
     1 - \sigma_{ji}(\mathbf{r}), & \text{if } i > j,
    \end{cases}
\end{equation}
which guarantees $\sigma_{ij}(\cdot) + \sigma_{ji}(\cdot) = 1$.
}

{
The shaping variables $\mathbf r$ provide a continuous description of how each pairwise overlap is distributed between its two modules. The complementary relation $\sigma_{ij}(\mathbf r)+\sigma_{ji}(\mathbf r)=1$ ensures that the complete overlap region is assigned without introducing an additional equality constraint. This formulation allows the overlap allocation to be optimized jointly with the bounding box variables.
}

\textbf{Definition 3}~(Rectilinear Polygon Representation).
{
\textit{The movable module set $\mathcal V_m$ is represented by the bounding box variables $\mathbf v$ and shaping variables $\mathbf r$ as
\begin{equation}
\begin{aligned}
\mathcal V_m(\mathbf v,\mathbf r)
:=\bigl\{v_i=\{b_i(\mathbf v),\;&
\{\sigma_{ij}(\mathbf r)\mid
\substack{v_j\in\mathcal V_m\\j\ne i}\}\}\\
&\bigm|\;v_i\in\mathcal V_m\bigr\}.
\end{aligned}
\label{eq:rectilinear-representation}
\end{equation}
}
Optimizing $\mathbf v$ and $\mathbf r$ jointly determines the movable bounding boxes $\mathbf b(\mathbf v)$ and their shaping proportions, enabling simultaneous optimization of module placement and rectilinear shapes.
}

\subsubsection{Pairwise Overlap Model}
To quantify the overlap between modules, we develop a differentiable pairwise overlap model based on the bounding boxes $\mathbf b(\mathbf v)$ recovered from the bounding box variables $\mathbf v$.

For two bounding boxes $b_i$ and $b_j$, the horizontal overlap is defined as:
{
\begin{equation}
    \nonumber
    o_{ij}^x(\mathbf b(\mathbf v)) :=
    \begin{cases}
        \beta_{ij} - \alpha_{ij}, & |d_x| \leq \alpha_{ij}, \\[6pt]
        \beta_{ij} - |d_x|, & \alpha_{ij} \leq |d_x| \leq \beta_{ij}, \\[6pt]
        0, & \beta_{ij} \leq |d_x|,
    \end{cases}
    \label{eq:overlap-piecewise}
\end{equation}}where $d_x = x_i - x_j$ denotes the horizontal distance between the module centers,
$\alpha_{ij} = \tfrac{|w_i - w_j| + \epsilon}{2}$ is the threshold at which the smaller module becomes fully contained within the larger one, and $\epsilon > 0$ is a small constant added to prevent numerical instability when $w_i = w_j$.
$\beta_{ij} = \tfrac{w_i + w_j}{2}$ defines the threshold beyond which no overlap occurs.
The three cases correspond to complete inclusion, partial overlap, and no overlap, respectively.
The vertical overlap $o_{ij}^y(\mathbf b(\mathbf v))$ is defined analogously.

\begin{figure}[htbp]
\centering
\resizebox{\hsize}{!}{
\begin{tikzpicture}[>=stealth]

\begin{axis}[
    width= 6cm,
    height=3.6cm,
    axis lines=middle,
    at = {(,0)},
    xlabel={$d_x$},
    ylabel={$\!o^{x}_{ij}$},
    xtick={-2,-1,0,1,2},
    xticklabels={$-\beta_{ij}$,$-\alpha_{ij}$,0,$\alpha_{ij}$,$\beta_{ij}$},
    ytick=\empty,
    xmin=-2.3, xmax=2.4,
    ymin=-0.1, ymax=2,
    samples=200,
    domain=-2:2,
    axis line style={->},
    clip=false
]
    \addplot[very thick, red!70!black, domain=-2:-1] {2 - abs(x)};
    \addplot[very thick, red!70!black, domain=-1:1] {2 - 1};
    \addplot[very thick, red!70!black, domain=1:2] {2 - abs(x)};
    \addplot[very thick, red!70!black, domain=-2.3:-2] {0};
    \addplot[very thick, red!70!black, domain=2:2.3] {0};
    \addplot[very thin, gray, dashed] coordinates {(-1,0) (-1,2)};
    \addplot[very thin, gray, dashed] coordinates {(1,0) (1,2)};
\end{axis}

\begin{axis}[
    width= 6cm,
    height=3.6cm,
    at={(4.8cm,0)},
    axis lines=middle,
    xlabel={$d_x$},
    ylabel={$\!\hat{o}^{x}_{ij}$},
    xtick={-2,-1,0,1,2},
    xticklabels={$-\beta_{ij}$,$-\alpha_{ij}$,0,$\alpha_{ij}$,$\beta_{ij}$},
    ytick=\empty,
    xmin=-2.3, xmax=2.4,
    ymin=-0.1, ymax=2,
    samples=200,
    axis line style={->},
    clip=false
]
    \addplot[very thick, red!70!black, domain=-2:-1] {2 - abs(x)};
    \addplot[very thick, red!70!black, domain=-1:1] {2 - (x^2 + 1)/2};
    \addplot[very thick, red!70!black, domain=1:2] {2 - abs(x)};
    \addplot[very thick, red!70!black, domain=-2.3:-2] {0};
    \addplot[very thick, red!70!black, domain=2:2.3] {0};
    \addplot[very thin, gray, dashed] coordinates {(-1,0) (-1,2)};
    \addplot[very thin, gray, dashed] coordinates {(1,0) (1,2)};
\end{axis}

\draw (1.95,-0.3) node [anchor=north west][inner sep=0.75pt]   [align=left][font=\large]  {(a)};
\draw (6.76,-0.3) node [anchor=north west][inner sep=0.75pt]   [align=left][font=\large]  {(b)};
\end{tikzpicture}
}
\caption{Illustration of the overlap function. (a) Original overlap function; (b) Smoothed overlap function.}
\label{fig:overlap_fn}
\end{figure}
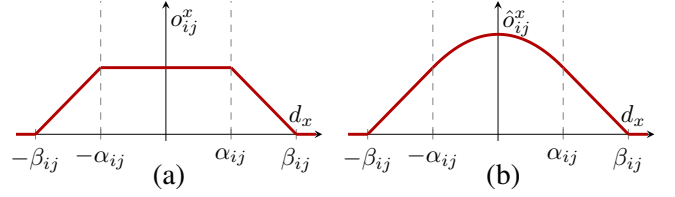

Although $o_{ij}^x(\cdot)$ accurately captures the horizontal overlap, it exhibits a limitation:
in the first case ($|d_x| \le \alpha_{ij}$), the partial derivative with respect to $d_x$ is zero, producing a constant plateau that impedes gradient-based optimization.
Without a nonzero gradient, smaller modules may remain nested inside larger ones, thereby introducing holes and violating the simplicity constraint~\eqref{eq:obj1:rectilinear}.

To address this issue, we propose a smoothed approximation:
{
\begin{equation}
    \nonumber
    \hat{o}_{ij}^x(\mathbf b(\mathbf v)) :=
    \begin{cases}
        \beta_{ij} - \tfrac{d_x^2 + \alpha_{ij}^2}{2\alpha_{ij}}, & |d_x| \leq \alpha_{ij}, \\[6pt]
        \beta_{ij} - |d_x|, & \alpha_{ij} \leq |d_x| \leq \beta_{ij}, \\[6pt]
        0, & \beta_{ij} \leq |d_x|,
    \end{cases}
    \label{eq:overlap-smooth}
\end{equation}
}
Here, the constant plateau is replaced by a quadratic curve, which provides nontrivial gradients and guarantees $C^1$ continuity, as illustrated in Figure~\ref{fig:overlap_fn}.
The vertical overlap $\hat{o}_{ij}^y(\mathbf b(\mathbf v))$ is defined analogously.

Finally, for two bounding boxes $b_i$ and $b_j$, the total overlap area is computed as the product of their horizontal and vertical overlaps:
{
\begin{equation}
    \nonumber
    o_{ij}(\mathbf b(\mathbf v))
    :=\hat{o}_{ij}^x(\mathbf b(\mathbf v))\,
    \hat{o}_{ij}^y(\mathbf b(\mathbf v)).
\end{equation}
}

\subsubsection{Differentiable Polygon Shaping Model}
While the pairwise overlap model $o_{ij}(\cdot)$ effectively quantifies the geometric intersection between rectangular bounding boxes $b_i$ and $b_j$, it inherently simplifies the underlying geometry by disregarding the possibility of \textit{polygon shaping}. The strength of our approach lies in the introduction of the shaping variables $\mathbf{r}$, which allows modules to dynamically deform and reassign portions of their shared overlap area, thereby enabling ``soft'' geometric behavior.

We extend the base overlap formulation into a differentiable polygon shaping model, allowing the joint optimization of both the bounding box variables $\mathbf v$ and shaping variables $\mathbf r$. This model explicitly connects the geometric overlap $o_{ij}(\mathbf b(\mathbf v))$ with the shaping proportion $\sigma_{ij}(\mathbf r)$.

Given the overlap model $o_{ij}(\cdot)$ and shaping proportion function $\sigma_{ij}(\cdot)$, the product $o_{ij}(\mathbf b(\mathbf v))\sigma_{ij}(\mathbf r)$ denotes the portion of the overlap area between modules $v_i$ and $v_j$ that is removed from module $v_i$ and assigned to $v_j$.
Accordingly, the total overlap area contributed by module $v_i$ is
{
\begin{equation}
    \mathcal O_i(\mathbf v,\mathbf r)
    :=\sum_{\substack{v_j\in\mathcal V_m\\j\ne i}}
    o_{ij}(\mathbf b(\mathbf v))\,
    \sigma_{ij}(\mathbf r)
    +\sum_{v_j\in\mathcal V_p}o_{ij}(\mathbf b),
    \label{eq:contributed-overlap-new}
\end{equation}
}
This quantity aggregates the overlap area donated by movable module $v_i$, weighted by the shaping proportion for other movable modules, while overlap with a preplaced module is assigned entirely to $v_i$. Because both $o_{ij}(\cdot)$ and $\sigma_{ij}(\cdot)$ are differentiable, the formulation enables gradient computation with respect to $\mathbf v$ and $\mathbf r$.

To satisfy the minimum area constraint~\eqref{eq:obj1:area}, the effective area remaining after polygon shaping must be no smaller than $A_i^{\min}$. Since $a_i$ is optimized jointly with the other bounding box variables, this condition is expressed as
{
\begin{equation}
    a_i-\mathcal O_i(\mathbf v,\mathbf r)
    \ge A_i^{\min}.
    \label{eq:variable-area-constraint}
\end{equation}
}
{
Optimizing $a_i$ allows the global floorplanning stage to select a suitable bounding box area for polygon shaping. Increasing $a_i$ provides more area for resolving overlap, while reducing it limits the geometric envelope and potential overlap. Its upper bound $A_i^{\min}/\mathrm{RR}_i^{\min}$ ensures consistency with the rectangle ratio constraint~\eqref{eq:obj1:rect-ratio}.
}

For improved numerical stability and scale invariance across modules of different areas, we define the normalized residual as
{
\begin{equation}
    \phi_i(\mathbf v,\mathbf r)
    :=\frac{\mathcal O_i(\mathbf v,\mathbf r)-a_i}
    {A_i^{\min}}+1\le0.
    \label{eq:new-normalized-residual}
\end{equation}
}
A positive residual indicates that the effective area of $v_i$ is smaller than $A_i^{\min}$, while a nonpositive residual indicates that the minimum area constraint is satisfied.

Overall, the differentiable shaping model enables joint optimization of module shape, area, and placement while incorporating the minimum area and rectangle ratio requirements into a continuous formulation suitable for global floorplanning.

\subsubsection{ALM-based Optimization Framework}
Given the rectilinear polygon representation, the pairwise overlap model, and the polygon shaping model, problem~\eqref{eq:obj1} can be reformulated as a constrained optimization problem over the bounding box variables $\mathbf v$ and shaping variables $\mathbf r$:
\begin{equation}
\label{eq:obj2}
\begin{aligned}
    \min_{\mathbf v\in\Omega,\,\mathbf r\in\mathbb R^{|\mathbf r|}}
    \quad &\sum_{e\in\mathcal E}\hat{\mathcal W}_e(\mathbf x,\mathbf y),\\
    \mathrm{s.t.}\quad
    &\boldsymbol{\phi}(\mathbf v,\mathbf r)\le\mathbf0.
\end{aligned}
\end{equation}
where $\hat{\mathcal W}_e(\cdot)$ denotes the smooth wirelength model defined in Eq.~\eqref{eq:wa-wirelength-model}. The fixed outline constraints~\eqref{eq:obj1:outline}, area bounds, and aspect ratio constraints~\eqref{eq:obj1:aspect} are enforced by projecting the updated variables onto the feasible region
\begin{equation}
\begin{aligned}
\Omega:=\bigl\{\mathbf v\bigm|\;&
A_i^{\min}\le a_i\le
\tfrac{A_i^{\min}}{\mathrm{RR}_i^{\min}},\quad v_i\in\mathcal V_s,\\
&\mathrm{AR}_i^l\le\mathrm{ar}_i\le\mathrm{AR}_i^u,
\quad v_i\in\mathcal V_s,\\
&(a_i,\mathrm{ar}_i)=(\bar a_i,\overline{\mathrm{ar}}_i),
\quad v_i\in\mathcal V_h,\\
&0\le x_i-\tfrac{w_i}{2}<x_i+\tfrac{w_i}{2}\le W,\\
&0\le y_i-\tfrac{h_i}{2}<y_i+\tfrac{h_i}{2}\le H,
\quad v_i\in\mathcal V_m\bigr\}.
\end{aligned}
\label{eq:new-feasible-region}
\end{equation}
Here, $w_i$ and $h_i$ are recovered by Eq.~\eqref{eq:area-ar-conversion}. The remaining constraints are represented by the normalized shaping-constraint residual vector $\boldsymbol{\phi}(\mathbf v,\mathbf r):=\{\phi_i(\mathbf v,\mathbf r)\mid v_i\in\mathcal V_m\}$.

To solve this constrained problem, we adopt the Augmented Lagrangian Method (ALM), which combines penalty and multiplier methods. In contrast to penalty methods in previous works~\cite{2024jigsawplanner,2024NTUFP,2006Analytical,2014F-FM,2022PeF,2023rotationfp}, which require the penalty parameter to approach infinity, ALM introduces Lagrange multipliers with finite penalty factors, enabling simultaneous control over feasibility and convergence~\cite{nocedal2006numerical}.

According to the ALM~\cite{nocedal2006numerical}, problem~\eqref{eq:obj2} can be transformed into an augmented Lagrangian function for the inequality constraints as follows:
\begin{equation}
\begin{aligned}
\mathcal L_\mu(\mathbf v,\mathbf r,\boldsymbol\lambda)
={}&\sum_{e\in\mathcal E}\hat{\mathcal W}_e(\mathbf x,\mathbf y)\\
&+\frac{\mu}{2}\left\|
\max\left\{\frac{\boldsymbol\lambda}{\mu}
+\boldsymbol{\phi}(\mathbf v,\mathbf r),\mathbf0\right\}
\right\|_2^2\\
&-\frac{\mu}{2}
\left\|\frac{\boldsymbol\lambda}{\mu}\right\|_2^2.
\end{aligned}
\label{eq:new-augmented-lagrangian}
\end{equation}
where $\boldsymbol\lambda=\{\lambda_i\ge0\mid v_i\in\mathcal V_m\}$ are the Lagrange multipliers and $\mu>0$ is the penalty parameter.

To quantitatively evaluate constraint satisfaction, the normalized primal infeasibility measure is defined as
\begin{equation}
\psi_k:=\left\|
\max\left\{\boldsymbol{\phi}(\mathbf v^{k+1},\mathbf r^{k+1}),
-\frac{\boldsymbol\lambda^k}{\mu_k}\right\}
\right\|_2.
\label{eq:new-alm-residual}
\end{equation}
which satisfies $\psi_k=0$ if and only if all constraints are satisfied. This measure directly guides the adaptive penalty update strategy and convergence control.

\begin{algorithm}[t]
\caption{ALM-based Global Floorplanning}
\label{alg:alm}
\begin{algorithmic}[1]
\Require Initial solution $\{\mathbf v^0,\mathbf r^0\}$, multipliers $\boldsymbol\lambda^0\geq0$, penalty parameter $\mu_0>0$, penalty factor $\tau>1$, scaling factor $0<\theta<1$, tolerance $\varepsilon_0=\theta\psi_0$, feasibility tolerance $\varepsilon_t\geq0$.
\Ensure Optimized solution $(\mathbf v^*,\mathbf r^*)$.
\For{$k=0,1,\ldots,N-1$}
    \State Initialize $(\mathbf v^{k,0},\mathbf r^{k,0})\gets(\mathbf v^k,\mathbf r^k)$ and restart the Adam state $\mathbf s^{k,0}$.
    \For{$t=0,1,\ldots,N_{\mathrm{in}}-1$} \Comment{Primal step}
        \State $\mathbf g^{k,t}\gets\nabla_{(\mathbf v,\mathbf r)}\mathcal L_{\mu_k}(\mathbf v^{k,t},\mathbf r^{k,t},\boldsymbol\lambda^k)$.
        \State $(\widetilde{\mathbf v},\mathbf r,\mathbf s)^{k,t+1}\gets\operatorname{Adam}((\mathbf v,\mathbf r)^{k,t},\mathbf g^{k,t},\mathbf s^{k,t})$.
        \State $\mathbf v^{k,t+1}\gets\Pi_{\Omega}(\widetilde{\mathbf v}^{k,t+1})$.
    \EndFor
    \State $(\mathbf v^{k+1},\mathbf r^{k+1})\gets(\mathbf v^{k,N_{\mathrm{in}}},\mathbf r^{k,N_{\mathrm{in}}})$.
    \State Evaluate $\psi_k$ using Eq.~\eqref{eq:new-alm-residual}.
    \If{$\psi_k<\varepsilon_k$} \Comment{Multiplier update}
        \State $\boldsymbol\lambda^{k+1}=\max\{0,\boldsymbol\lambda^k+\mu_k\boldsymbol{\phi}(\mathbf v^{k+1},\mathbf r^{k+1})\}$,
        \State $\mu_{k+1}=\mu_k$, $\varepsilon_{k+1}=\theta\psi_k$,
    \Else \Comment{Penalty update}
        \State $\boldsymbol\lambda^{k+1}=\boldsymbol\lambda^k$,
        \State $\mu_{k+1}=\tau\mu_k$, $\varepsilon_{k+1}=\varepsilon_k$.
    \EndIf
    \If{$\psi_k\leq\varepsilon_t$} \Comment{Stopping criterion}
        \State \Return $\{\mathbf v^{k+1},\mathbf r^{k+1}\}$.
    \EndIf
\EndFor
\State \Return $\{\mathbf v^N,\mathbf r^N\}$.
\end{algorithmic}
\end{algorithm}

The complete ALM-based global floorplanning algorithm is summarized in Algorithm~\ref{alg:alm}. Lines 1--8 approximately solve each primal subproblem. The bounding box and shaping variables are first initialized from the current outer iterate, after which their joint gradient is evaluated and used by Adam~\cite{adam2014method}. The updated bounding box variables are projected onto $\Omega$, and the resulting primal solution is used to evaluate $\psi_k$. Lines 9--16 then perform the dual step. When $\psi_k$ is below the adaptive tolerance $\varepsilon_k$, the reduction in constraint violation is sufficient to update the multipliers while retaining the current penalty parameter. Otherwise, the multipliers are retained and the penalty parameter is increased to strengthen constraint enforcement.
This adaptive strategy avoids increasing the penalty unnecessarily while preventing multiplier updates before adequate progress has been made in the primal subproblem. The algorithm terminates when the target feasibility tolerance is reached or the maximum number of outer iterations is exhausted.
\begin{figure}[t]
\centering

\tikzset{every picture/.style={line width=0.75pt}}

\begin{tikzpicture}[x=0.75pt,y=0.75pt,yscale=-1.1,xscale=1.1]

\draw  [line width=0.75]  (1,1) -- (131,1) -- (131,131) -- (1,131) -- cycle ;
\draw  [fill={rgb, 255:red, 190; green, 233; blue, 232 }  ,fill opacity=0.8 ][line width=0.75]  (1,1) -- (66,1) -- (66,76) -- (1,76) -- cycle ;
\draw  [fill={rgb, 255:red, 142; green, 219; blue, 229 }  ,fill opacity=0.8 ][line width=0.75]  (41,1) -- (101,1) -- (101,51) -- (41,51) -- cycle ;
\draw  [fill={rgb, 255:red, 99; green, 178; blue, 224 }  ,fill opacity=0.8 ][line width=0.75]  (101,1) -- (131,1) -- (131,51) -- (101,51) -- cycle ;
\draw  [fill={rgb, 255:red, 95; green, 168; blue, 211 }  ,fill opacity=0.8 ][line width=0.75]  (1,61) -- (66,61) -- (66,131) -- (1,131) -- cycle ;
\draw  [fill={rgb, 255:red, 202; green, 233; blue, 255 }  ,fill opacity=0.8 ][line width=0.75]  (66,36) -- (131,36) -- (131,96) -- (66,96) -- cycle ;
\draw  [fill={rgb, 255:red, 98; green, 182; blue, 203 }  ,fill opacity=0.8 ][line width=0.75]  (56,81) -- (131,81) -- (131,131) -- (56,131) -- cycle ;
\draw  [fill={rgb, 255:red, 190; green, 233; blue, 232 }  ,fill opacity=0.8 ][line width=0.75]  (211,1) -- (211,21) -- (236,21) -- (236,76) -- (201,76) -- (201,61) -- (171,61) -- (171,1) -- cycle ;
\draw  [fill={rgb, 255:red, 142; green, 219; blue, 229 }  ,fill opacity=0.8 ][line width=0.75]  (211,1) -- (271,1) -- (271,36) -- (256,36) -- (256,51) -- (236,51) -- (236,21) -- (211,21) -- cycle ;
\draw  [fill={rgb, 255:red, 99; green, 178; blue, 224 }  ,fill opacity=0.8 ][line width=0.75]  (271,1) -- (301,1) -- (301,51) -- (286,51) -- (286,36) -- (271,36) -- cycle ;
\draw  [fill={rgb, 255:red, 202; green, 233; blue, 255 }  ,fill opacity=0.8 ][line width=0.75]  (286,36) -- (286,51) -- (301,51) -- (301,96) -- (286,96) -- (286,81) -- (236,81) -- (236,51) -- (256,51) -- (256,36) -- cycle ;
\draw  [fill={rgb, 255:red, 98; green, 182; blue, 203 }  ,fill opacity=0.8 ][line width=0.75]  (286,81) -- (286,96) -- (301,96) -- (301,131) -- (226,131) -- (226,116) -- (236,116) -- (236,81) -- cycle ;
\draw  [fill={rgb, 255:red, 95; green, 168; blue, 211 }  ,fill opacity=0.8 ][line width=0.75]  (171,61) -- (201,61) -- (201,76) -- (236,76) -- (236,116) -- (226,116) -- (226,131) -- (171,131) -- cycle ;
\draw [line width=0.75]  [dash pattern={on 4.5pt off 4.5pt}]  (31,56) -- (31,85.25) ;
\draw [line width=0.75]  [dash pattern={on 4.5pt off 4.5pt}]  (116,73.4) -- (116,102.65) ;
\draw [line width=0.75]  [dash pattern={on 4.5pt off 4.5pt}]  (116,31) -- (116,60.25) ;
\draw [line width=0.75]  [dash pattern={on 4.5pt off 4.5pt}]  (46.5,116) -- (75.5,116.05) ;
\draw [line width=0.75]  [dash pattern={on 4.5pt off 4.5pt}]  (34,21) -- (76,20.95) ;
\draw [line width=0.75]  [dash pattern={on 4.5pt off 4.5pt}]  (86,31) -- (86,60.25) ;
\draw  [line width=0.75]  (171,1) -- (301,1) -- (301,131) -- (171,131) -- cycle ;

\draw (59,135) node [anchor=north west][inner sep=0.75pt]   [align=left][font=\large]  {(a)};
\draw (229,135) node [anchor=north west][inner sep=0.75pt]   [align=left] [font=\large]{(b)};
\end{tikzpicture}
\caption{Illustration of shape determination. (a) Overlapping bounding boxes;~(b) Result of polygon shape determination.}
\label{fig:polygonclip}
\end{figure}
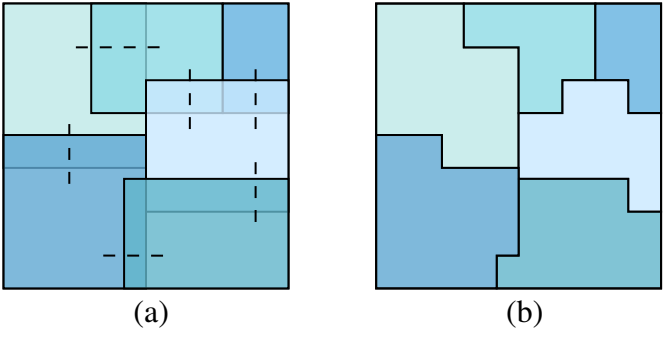

\subsubsection{Polygon Shape Determination}
Given the optimized variables $(\mathbf v^*,\mathbf r^*)$ returned by the ALM framework, we obtain the final bounding box solution $\mathbf b^*=\mathbf b(\mathbf v^*)$ and the shaping solution $\mathbf r^*$. Although the continuous optimization successfully minimizes wirelength and ensures feasibility $(\psi^* < \varepsilon_t)$, the result still represents an abstract configuration with overlapping rectangular bounding boxes rather than explicit polygonal shapes.
This step converts the abstract solution into actual rectilinear polygons by geometrically resolving overlaps according to the optimized shaping proportions $\sigma_{ij}(\mathbf{r}^*)$, producing a polygon module set $\mathcal{P}$ defined as:
$$
\mathcal{P} := \{\, p_i = \{(x_1, y_1), (x_2, y_2), \ldots, (x_n, y_n)\} \mid i \in m\},
$$
where each $p_i$ represents the feasible shape of module $i$ after overlap resolution. The procedure iterates over all pairs of overlapping bounding boxes and performs geometric cuts and assignments, as illustrated in Figure~\ref{fig:polygonclip}.

For each overlapping pair $(b_i, b_j)$, the overlapping region $O_{ij}$ is computed as the intersection of the two rectangles. Since both boxes are rectangular, $O_{ij}$ is also rectangular with center $(x_{ij}, y_{ij})$, width $w_{ij}$, and height $h_{ij}$.

To partition and assign $O_{ij}$, we must use either a single $x$-cut (vertical cut) or $y$-cut (horizontal cut) to maintain the rectilinear geometry of the resulting polygons. The decision of cutting direction depends on the relative dimensions of $O_{ij}$ with respect to $b_i$ and $b_j$.
We calculate the normalized maximum dimension ratios $\rho_x$ and $\rho_y$ as follows:
$$
\rho_x = \max\left(\frac{w_{ij}}{w_i}, \frac{w_{ij}}{w_j}\right), \quad
\rho_y = \max\left(\frac{h_{ij}}{h_i}, \frac{h_{ij}}{h_j}\right),
$$
and select a $y$-cut if $\rho_y > \rho_x$; otherwise, an $x$-cut is used.

After determining the cutting direction, $O_{ij}$ is partitioned and assigned based on the optimized shaping proportions $\sigma_{ij}(\mathbf{r}^*)$ (where $\sigma_{ij} + \sigma_{ji} = 1$) and relative module coordinates. Taking the $y$-cut as an illustrative example, if $y_i \le y_j$ (module $i$ is below module $j$), the overlap region is divided horizontally.
The rectangular region $O_{ij}^i$ allocated to $v_i$ is:
$$
O_{ij}^i = \{x_{ij}, y_{ij}-\frac{\sigma_{ij}(\mathbf{r}^*)h_{ij}}{2}, w_{ij}, \sigma_{ji}(\mathbf{r}^*) h_{ij}\},
$$
while the remaining region $O_{ij} - O_{ij}^i$ is assigned to $v_j$.

This approach ensures that the cut aligns with the dominant dimension of $O_{ij}$ and allocates regions according to relative positions, thereby preserving spatial consistency and avoiding violation of the simplicity constraint~\eqref{eq:obj1:rectilinear}.

After all overlapping module pairs have been processed, the hard and preplaced rectangles are subtracted from the soft polygons. Since these Boolean operations may introduce violations of the minimum inner width constraint~\eqref{eq:obj1:inner-width}, we further enforce this constraint before outputting the polygon shapes. Each soft polygon is sliced along both the horizontal and vertical directions. A slice whose thickness is below $\omega^{\min}$ is identified as a violation. The slice is removed only when the subtraction keeps the remaining polygon connected, preserves positive area, and reduces the number of polygon edges. This procedure is repeated until no removable violation remains. The resulting polygons form the final polygon module set $\mathcal P$ and are passed to legalization.

\subsection{Legalization}\label{sec:legalization}

After obtaining a global floorplanning solution, certain modules may still violate the minimum area constraint~\eqref{eq:obj1:area} due to limited local area availability.
Therefore, a legalization step is necessary to resolve these violations and ensure a valid floorplan.
Previous methods~\cite{2023tofu, 2024jigsawplanner, 2024NTUFP} employ heuristic approaches to address this issue,
but such methods lack theoretical guarantees for obtaining valid solutions.
\subsubsection{MCF-based Area Redistribution}
To address this problem, we propose a minimum-cost flow (MCF)-based area redistribution method that redistributes excessive or insufficient area  over the contact graph until all modules meet the minimum area constraint.

\input{include/Figure/graph.tex}

Given a polygon module set $\mathcal{P}$, the contact graph $\mathcal G_{\mathrm C}=(\mathcal N,\mathcal A)$ is constructed, where each polygon $p_i \in \mathcal{P}$ is a node $n_i \in \mathcal{N}$. A set of arcs $a_{ij} = \{a_{ij}^k \mid k \in K\}$ is added if $p_i$ and $p_j$ share a set of boundary segments.  This segment set is defined solely by the longer of the shared horizontal or vertical segment sets to avoid expanding mutual exclusion relationships, as shown in Figure~\ref{fig:graph}(a).
Nodes $n_i \in \mathcal{N}_s$ with area deficiency (i.e., $\text{Area}(p_i) < A_i^{\min}$) act as \textit{sources}, while nodes $n_i \in \mathcal{N}_t$ with redundant area (i.e., $\text{Area}(p_i) > A_i^{\min}$) or whitespace act as \textit{sinks}.
Area redistribution occurs along the arcs of this graph, representing the local geometric transfer of space between neighboring modules.

To maintain the quality of final solutions, we aim to minimize the total Manhattan displacement relative to the global floorplanning solution during area redistribution.
Therefore, we formulate the area transfer problem as follows:

\begin{equation}
    \label{eq:obj3}
    \begin{aligned}
        \min_{\mathcal{T}\in \mathbb{R}^{|\mathcal{A}|}} \quad & \!\!\! \sum_{a_{ij}^k \in \mathcal{A}} \!\! \mathcal{I}_{ij}^k \mathcal{T}_{ij}^k \\
        \text{s.t.}\quad\quad
        & \mathcal{T}_{ij}^k \le \mathcal{C}_{ij}^k, \quad \forall a_{ij}^k \in \mathcal{A}, \\
        & \mathcal{T}_{ij}^k + \mathcal{T}_{ji}^k = 0, \quad \forall a_{ij}^k, a_{ji}^k \in \mathcal{A}, \\
        &\!\!\!\! \sum_{j:a_{ij}^k \in \mathcal{A}} \!\! \mathcal{T}_{ij}^k = 0, \quad \forall n_i \notin  \mathcal{N}_s \cup  \mathcal{N}_t, \\
        &\!\!\!\!\sum_{j:a_{ij}^k \in \mathcal{A}} \!\! \mathcal{T}_{ij}^k = \mathcal{D}, \quad \forall n_i \in \mathcal{N}_s, \\
        &\!\!\!\!\sum_{j:a_{ij}^k \in \mathcal{A}} \!\! \mathcal{T}_{ij}^k = -\mathcal{D}, \quad \forall n_i \in \mathcal{N}_t,
    \end{aligned}
\end{equation}
where $\mathcal{T}_{ij}^k$ represents area transferred along arc $a_{ij}^k$ from module $p_j$ to $p_i$.
The $\mathcal{I}_{ij}^k$ represents the unit cost of area transferred along arc $a_{ij}^k$, modeling the Manhattan displacement incurred by area transfer.
As shown in Figure~\ref{fig:graph}(b), $\mathcal{I}_{ij}^{\delta}$ is defined as:
\begin{equation}
\nonumber
\mathcal{I}_{ij}^{\delta} :=
\begin{cases}
\frac{1}{2|\delta|}, & \delta \cap b_i \neq \emptyset, \\
0, & \text{otherwise}.
\end{cases}
\end{equation}
Here, if $\delta$ intersects with $b_i$, then extending $\delta$ will cause a Manhattan displacement of $\mathcal{T}_{ij}^{\delta}/(2|\delta|)$ to the center of $b_i$. Therefore, the unit transfer cost is set to $1/(2|\delta|)$. Conversely, if $\delta$ does not intersect with $b_i$, transferring area along this boundary will not incur any displacement, and thus the unit transfer cost is set to zero.
The objective is to minimize the product $\mathcal{I}_{ij}^k \mathcal{T}_{ij}^k$, which therefore quantifies the Manhattan displacement incurred by transferring area along that arc $a_{ij}^k$.

The first constraint limits the transfer area along each arc by its transfer capacity $\mathcal{C}_{ij}^k$. The transfer capacity determines the maximum transferable area without violating aspect ratio or rectangular ratio constraints, as illustrated in Figure~\ref{fig:graph}(b), the transfer capacity of arc $a_{ij}^{\delta}$ is defined as:
\begin{equation}
\nonumber
\mathcal{C}_{ij}^{\delta} :=
\begin{cases}
|\delta| \cdot \min\{\delta_i, \delta_j\}, & \delta \cap b_i \neq \emptyset, \\[4pt]
\text{Area}(b_i \cap p_j) , & \text{otherwise},
\end{cases}
\end{equation}
where $\delta_i$ and $\delta_j$ denote the maximum feasible boundary displacement of $p_i$ and $p_j$ along $\delta$, respectively.
For shared boundaries that do not intersect with the bounding boxes, the transfer capacity is determined by the available overlapping area between the two bounding boxes.
The second constraint enforces antisymmetric demand consistency between opposite arcs.
The third constraint ensures area balance for all internal modules.
The total area demand $\mathcal{D}$ is defined as:
$$\mathcal{D} = \sum_{i}^m\max\{A_i^{min}-\text{Area}(p_i),0\}.$$
which represents the required area to satisfy the minimum area constraint.
The fourth and fifth constraints specify area demand and supply conditions, where source modules require an inflow of $\mathcal{D}$ area, and sink modules provide an equivalent outflow.

The problem~\eqref{eq:obj3} is a classical MCF problem.
It can be efficiently solved using the network simplex algorithm~\cite{kelly1991minimum}, which guarantees an optimal flow distribution that satisfies all capacity and conservation constraints. After obtaining the optimal solution $\mathcal T^*$, each $\mathcal T_{ij}^{k*}>0$ represents the polygon level area transferred from donor $p_j$ to receiver $p_i$ through the geometric region associated with arc $a_{ij}^k$. These transfer amounts are passed to the rectangular decomposition stage, where they are attributed to the subrectangles of the corresponding polygons.

\subsubsection{Rectangular Decomposition}
{
The polygon level flow cannot be applied directly to the subsequent geometric optimization because a rectilinear polygon may contain multiple shape components with different movement directions. We therefore decompose each soft polygon $p_i$ into a set of connected rectangles and assign a target area to each resulting rectangle.

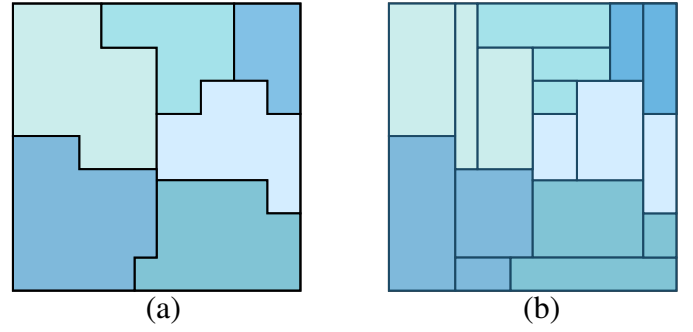
\begin{figure}[t]
\centering
\color{black}

\tikzset{every picture/.style={line width=0.75pt}}

\begin{tikzpicture}[x=0.75pt,y=0.75pt,yscale=-0.74,xscale=0.74]

\draw  [color={rgb, 255:red, 27; green, 73; blue, 101 }  ,draw opacity=1 ][fill={rgb, 255:red, 99; green, 178; blue, 224 }  ,fill opacity=0.8 ][line width=0.75]  (406,1) -- (451,1) -- (451,75.9) -- (428.5,75.9) -- (428.5,53.43) -- (406,53.43) -- cycle ;
\draw  [color={rgb, 255:red, 0; green, 0; blue, 0 }  ,draw opacity=1 ][line width=0.75]  (256,1) -- (451,1) -- (451,195.75) -- (256,195.75) -- cycle ;
\draw  [color={rgb, 255:red, 27; green, 73; blue, 101 }  ,draw opacity=1 ][fill={rgb, 255:red, 95; green, 168; blue, 211 }  ,fill opacity=0.8 ][line width=0.75]  (256,90.88) -- (301,90.88) -- (301,195.75) -- (256,195.75) -- cycle ;
\draw  [color={rgb, 255:red, 27; green, 73; blue, 101 }  ,draw opacity=1 ][fill={rgb, 255:red, 95; green, 168; blue, 211 }  ,fill opacity=0.8 ][line width=0.75]  (301,113.36) -- (353.5,113.36) -- (353.5,173.28) -- (301,173.28) -- cycle ;
\draw  [color={rgb, 255:red, 27; green, 73; blue, 101 }  ,draw opacity=1 ][fill={rgb, 255:red, 95; green, 168; blue, 211 }  ,fill opacity=0.8 ][line width=0.75]  (301,173.28) -- (338.5,173.28) -- (338.5,195.75) -- (301,195.75) -- cycle ;
\draw  [color={rgb, 255:red, 27; green, 73; blue, 101 }  ,draw opacity=1 ][fill={rgb, 255:red, 98; green, 182; blue, 203 }  ,fill opacity=0.8 ][line width=0.75]  (353.5,120.85) -- (428.5,120.85) -- (428.5,173.28) -- (353.5,173.28) -- cycle ;
\draw  [color={rgb, 255:red, 27; green, 73; blue, 101 }  ,draw opacity=1 ][fill={rgb, 255:red, 98; green, 182; blue, 203 }  ,fill opacity=0.8 ][line width=0.75]  (338.5,173.28) -- (451,173.28) -- (451,195.75) -- (338.5,195.75) -- cycle ;
\draw  [color={rgb, 255:red, 27; green, 73; blue, 101 }  ,draw opacity=1 ][fill={rgb, 255:red, 98; green, 182; blue, 203 }  ,fill opacity=0.8 ][line width=0.75]  (428.5,143.32) -- (451,143.32) -- (451,173.28) -- (428.5,173.28) -- cycle ;
\draw  [color={rgb, 255:red, 27; green, 73; blue, 101 }  ,draw opacity=1 ][fill={rgb, 255:red, 202; green, 233; blue, 255 }  ,fill opacity=0.8 ][line width=0.75]  (383.5,53.43) -- (428.5,53.43) -- (428.5,120.85) -- (383.5,120.85) -- cycle ;
\draw  [color={rgb, 255:red, 27; green, 73; blue, 101 }  ,draw opacity=1 ][fill={rgb, 255:red, 202; green, 233; blue, 255 }  ,fill opacity=0.8 ][line width=0.75]  (428.5,75.9) -- (451,75.9) -- (451,143.32) -- (428.5,143.32) -- cycle ;
\draw  [color={rgb, 255:red, 27; green, 73; blue, 101 }  ,draw opacity=1 ][fill={rgb, 255:red, 202; green, 233; blue, 255 }  ,fill opacity=0.8 ][line width=0.75]  (353.5,75.9) -- (383.5,75.9) -- (383.5,120.85) -- (353.5,120.85) -- cycle ;
\draw  [color={rgb, 255:red, 27; green, 73; blue, 101 }  ,draw opacity=1 ][fill={rgb, 255:red, 142; green, 219; blue, 229 }  ,fill opacity=0.8 ][line width=0.75]  (316,1) -- (406,1) -- (406,30.96) -- (316,30.96) -- cycle ;
\draw  [color={rgb, 255:red, 27; green, 73; blue, 101 }  ,draw opacity=1 ][fill={rgb, 255:red, 142; green, 219; blue, 229 }  ,fill opacity=0.8 ][line width=0.75]  (353.5,30.96) -- (406,30.96) -- (406,53.43) -- (353.5,53.43) -- cycle ;
\draw  [color={rgb, 255:red, 27; green, 73; blue, 101 }  ,draw opacity=1 ][fill={rgb, 255:red, 142; green, 219; blue, 229 }  ,fill opacity=0.8 ][line width=0.75]  (353.5,53.43) -- (383.5,53.43) -- (383.5,75.9) -- (353.5,75.9) -- cycle ;
\draw  [color={rgb, 255:red, 27; green, 73; blue, 101 }  ,draw opacity=1 ][fill={rgb, 255:red, 99; green, 178; blue, 224 }  ,fill opacity=0.8 ][line width=0.75]  (406,1) -- (428.5,1) -- (428.5,53.43) -- (406,53.43) -- cycle ;
\draw  [color={rgb, 255:red, 27; green, 73; blue, 101 }  ,draw opacity=1 ][fill={rgb, 255:red, 99; green, 178; blue, 224 }  ,fill opacity=0.8 ][line width=0.75]  (428.5,1) -- (451,1) -- (451,75.9) -- (428.5,75.9) -- cycle ;
\draw  [color={rgb, 255:red, 27; green, 73; blue, 101 }  ,draw opacity=1 ][fill={rgb, 255:red, 190; green, 233; blue, 232 }  ,fill opacity=0.8 ][line width=0.75]  (256,1) -- (301,1) -- (301,90.88) -- (256,90.88) -- cycle ;
\draw  [color={rgb, 255:red, 27; green, 73; blue, 101 }  ,draw opacity=1 ][fill={rgb, 255:red, 190; green, 233; blue, 232 }  ,fill opacity=0.8 ][line width=0.75]  (316,30.96) -- (353.5,30.96) -- (353.5,113.36) -- (316,113.36) -- cycle ;
\draw  [color={rgb, 255:red, 27; green, 73; blue, 101 }  ,draw opacity=1 ][fill={rgb, 255:red, 190; green, 233; blue, 232 }  ,fill opacity=0.8 ][line width=0.75]  (301,1) -- (316,1) -- (316,113.36) -- (301,113.36) -- cycle ;
\draw  [color={rgb, 255:red, 0; green, 0; blue, 0 }  ,draw opacity=1 ][fill={rgb, 255:red, 190; green, 233; blue, 232 }  ,fill opacity=0.8 ][line width=0.75]  (61,1) -- (61,30.96) -- (98.5,30.96) -- (98.5,113.36) -- (46,113.36) -- (46,90.88) -- (1,90.88) -- (1,1) -- cycle ;
\draw  [color={rgb, 255:red, 0; green, 0; blue, 0 }  ,draw opacity=1 ][fill={rgb, 255:red, 142; green, 219; blue, 229 }  ,fill opacity=0.8 ][line width=0.75]  (61,1) -- (151,1) -- (151,53.43) -- (128.5,53.43) -- (128.5,75.9) -- (98.5,75.9) -- (98.5,30.96) -- (61,30.96) -- cycle ;
\draw  [color={rgb, 255:red, 0; green, 0; blue, 0 }  ,draw opacity=1 ][fill={rgb, 255:red, 99; green, 178; blue, 224 }  ,fill opacity=0.8 ][line width=0.75]  (151,1) -- (196,1) -- (196,75.9) -- (173.5,75.9) -- (173.5,53.43) -- (151,53.43) -- cycle ;
\draw  [color={rgb, 255:red, 0; green, 0; blue, 0 }  ,draw opacity=1 ][fill={rgb, 255:red, 202; green, 233; blue, 255 }  ,fill opacity=0.8 ][line width=0.75]  (173.5,53.43) -- (173.5,75.9) -- (196,75.9) -- (196,143.32) -- (173.5,143.32) -- (173.5,120.85) -- (98.5,120.85) -- (98.5,75.9) -- (128.5,75.9) -- (128.5,53.43) -- cycle ;
\draw  [color={rgb, 255:red, 0; green, 0; blue, 0 }  ,draw opacity=1 ][fill={rgb, 255:red, 98; green, 182; blue, 203 }  ,fill opacity=0.8 ][line width=0.75]  (173.5,120.85) -- (173.5,143.32) -- (196,143.32) -- (196,195.75) -- (83.5,195.75) -- (83.5,173.28) -- (98.5,173.28) -- (98.5,120.85) -- cycle ;
\draw  [color={rgb, 255:red, 0; green, 0; blue, 0 }  ,draw opacity=1 ][fill={rgb, 255:red, 95; green, 168; blue, 211 }  ,fill opacity=0.8 ][line width=0.75]  (1,90.88) -- (46,90.88) -- (46,113.36) -- (98.5,113.36) -- (98.5,173.28) -- (83.5,173.28) -- (83.5,195.75) -- (1,195.75) -- cycle ;
\draw  [color={rgb, 255:red, 0; green, 0; blue, 0 }  ,draw opacity=1 ][line width=0.75]  (1,1) -- (196,1) -- (196,195.75) -- (1,195.75) -- cycle ;

\draw (89,197.67) node [anchor=north west][inner sep=0.75pt]   [align=left][font=\large]  {(a)};
\draw (345,197.7) node [anchor=north west][inner sep=0.75pt]   [align=left] [font=\large]{(b)};
\end{tikzpicture}
\caption{Illustration of rectangular decomposition. (a) The original polygonal layout. (b) The decomposed rectangular layout.}
\label{fig:decomposition}
\end{figure}

As illustrated in Figure~\ref{fig:decomposition}(a), the decomposition is performed recursively. If $p_i$ is already rectangular, its bounding box is returned directly. Otherwise, horizontal and vertical rectangular slices are generated, and each cut that partitions these slices into two connected child polygons is examined. Among all valid cuts, we select the one that minimizes the area imbalance
\begin{equation}
    \left|
    \text{Area}\bigl(p_i^{(1)}\bigr)
    -\text{Area}\bigl(p_i^{(2)}\bigr)
    \right|,
    \label{eq:decomposition-balance}
\end{equation}
where $p_i^{(1)}$ and $p_i^{(2)}$ are the two child polygons. The same procedure is then applied to both children until every resulting component is rectangular. If no valid cut preserves the connectivity of both children, the bounding box of the current polygon is used as a fallback rectangle.

Let $\mathcal S_i$ denote the set of subrectangles obtained from $p_i$ and belonging to soft module $v_i$, and let $S_i^j\in\mathcal S_i$ denote one of its subrectangles. Each subrectangle is initialized with a target area equal to its geometric area
\begin{equation}
    A_i^{j,\mathrm{tar}}
    :=\text{Area}(S_i^j).
    \label{eq:initial-rectangle-target}
\end{equation}
The positive flows in $\mathcal T^*$ are subsequently attributed to these target areas. For a transfer $\mathcal T_{ij}^{k*}>0$, the geometric region associated with $a_{ij}^k$ is divided into one or more slices $Q$ on the donor side facing receiver $p_i$, as illustrated in Figure~\ref{fig:decomposition}(b). The slice areas $\text{Area}(Q)$ sum to $\mathcal T_{ij}^{k*}$.

Receiver $p_i$ gains $\text{Area}(Q)$, which is distributed among its subrectangles according to their shared boundary lengths with $Q$
\begin{equation}
    \Delta A_i^{j,\mathrm{tar}}
    =\text{Area}(Q)
    \frac{\ell(Q,S_i^j)}
    {\displaystyle\sum_{S_i^k\in\mathcal S_i}
    \ell(Q,S_i^k)},
    \label{eq:receiver-area-attribution}
\end{equation}
where $\ell(Q,S_i^j)$ denotes the length of their common boundary. If donor $p_j$ is a soft module, the same area is subtracted from its subrectangles in proportion to their overlap areas with $Q$
\begin{equation}
    \Delta A_j^{k,\mathrm{tar}}
    =-\text{Area}(Q)
    \frac{\text{Area}(S_j^k\cap Q)}
    {\displaystyle\sum_{S_j^l\in\mathcal S_j}
    \text{Area}(S_j^l\cap Q)}.
    \label{eq:donor-area-attribution}
\end{equation}
No donor target is reduced when the transferred area is supplied by whitespace.

After all positive flows have been attributed, each target area is limited to a small positive value to avoid degenerate rectangles. If the target areas of soft module $p_i$ still sum to less than $A_i^{\min}$, the remaining deficit is distributed in proportion to the current targets
\begin{equation}
\begin{aligned}
A_i^{j,\mathrm{tar}}\gets{}&A_i^{j,\mathrm{tar}}
+\frac{A_i^{j,\mathrm{tar}}}
{\displaystyle\sum_{S_i^k\in\mathcal S_i}A_i^{k,\mathrm{tar}}}
\left(A_i^{\min}
-\sum_{S_i^k\in\mathcal S_i}A_i^{k,\mathrm{tar}}\right).
\end{aligned}
\label{eq:rectangle-deficit-fill}
\end{equation}
The decomposition stage outputs the rectangle set
\begin{equation}
\mathcal R:=\left(\bigcup_{v_i\in\mathcal V_s}\mathcal S_i\right)
\cup\{b_i\mid v_i\in\mathcal V_h\cup\mathcal V_p\},
\label{eq:legalization-rectangle-set}
\end{equation}
together with the target areas $A_i^{j,\mathrm{tar}}$ of its soft subrectangles. These outputs are passed to constraint graph construction.
}

\subsubsection{Constraint Graph Construction}
{
The rectangles produced by decomposition, together with the hard and preplaced modules, are organized into the constraint graph set
\begin{equation}
\mathcal G
:=\left\{
\mathcal G_{\alpha}=(\mathcal N_{\alpha},\mathcal E_{\alpha})
\mathrel{\big|}
\alpha\in\{\mathrm H,\mathrm V\}
\right\},
\label{eq:constraint-graphs}
\end{equation}
where $\mathcal G_{\mathrm H}$ and $\mathcal G_{\mathrm V}$ are the horizontal constraint graph (HCG) and vertical constraint graph (VCG), respectively.
The node set consists of the rectangle nodes $\mathcal N_{\mathcal R}:=\{n_r\mid r\in\mathcal R\}$ and the source and sink nodes representing the two boundaries in each direction
\begin{equation}
\mathcal N_{\alpha}
=
\mathcal N_{\mathcal R}\cup\{s_{\alpha},t_{\alpha}\},
\qquad \alpha\in\{\mathrm H,\mathrm V\}.
\label{eq:constraint-graph-nodes}
\end{equation}
Each edge set contains three types of relations
\begin{equation}
\mathcal E_{\alpha}
=
\mathcal E_{\alpha}^{\mathrm I}
\mathbin{\dot\cup}
\mathcal E_{\alpha}^{=}
\mathbin{\dot\cup}
\mathcal E_{\alpha}^{\mathrm O},
\qquad \alpha\in\{\mathrm H,\mathrm V\},
\label{eq:typed-constraint-edges}
\end{equation}
where inequality edges $\mathcal E_{\alpha}^{\mathrm I}$ impose separation, equality edges $\mathcal E_{\alpha}^{=}$ impose exact adjacency, and overlap edges $\mathcal E_{\alpha}^{\mathrm O}$ maintain a positive overlap in the corresponding direction.

\input{include/Figure/constraint_graph}

To determine whether a rectangle pair should be separated horizontally or vertically, we compute the penetration depths from its geometry before legalization and select the direction by
\begin{equation}
\begin{aligned}
&\pi_{uv}^{x}:=\frac{w_u^0+w_v^0}{2}-|x_u^0-x_v^0|,\\
&\pi_{uv}^{y}:=\frac{h_u^0+h_v^0}{2}-|y_u^0-y_v^0|,\\
&\operatorname{dir}_{uv}:=
\begin{cases}
\mathrm H, & \pi_{uv}^{x}<\pi_{uv}^{y},\\
\mathrm V, & \text{otherwise}.
\end{cases}
\end{aligned}
\label{eq:constraint-direction-selection}
\end{equation}
An edge in $\mathcal G_{\mathrm H}$ is directed from the initially left rectangle to the right rectangle, while an edge in $\mathcal G_{\mathrm V}$ is directed from the initially bottom rectangle to the top rectangle. Rectangle indices break ties between equal coordinates.

For subrectangles of the same soft module, adjacency is determined from their original shared boundary. Define $x_r^{\pm}:=x_r^0\pm w_r^0/2$, $y_r^{\pm}:=y_r^0\pm h_r^0/2$, and $[z]_+:=\max\{z,0\}$. The shared-boundary lengths are
\begin{equation}
\begin{aligned}
&\ell_{uv}^{\mathrm H}>0
\Longleftrightarrow
\begin{cases}
x_u^+=x_v^-\ \text{or}\ x_v^+=x_u^-,\\
\min\{y_u^+,y_v^+\}>\max\{y_u^-,y_v^-\},
\end{cases}\\
&\ell_{uv}^{\mathrm V}>0
\Longleftrightarrow
\begin{cases}
y_u^+=y_v^-\ \text{or}\ y_v^+=y_u^-,\\
\min\{x_u^+,x_v^+\}>\max\{x_u^-,x_v^-\}.
\end{cases}
\end{aligned}
\label{eq:subrectangle-adjacency}
\end{equation}
Here, $\ell_{uv}^{\mathrm H}$ and $\ell_{uv}^{\mathrm V}$ denote the vertical and horizontal shared-boundary lengths, respectively. Thus, the first condition identifies a left-right adjacent pair, while the second identifies a bottom-top adjacent pair. The complete graph construction is summarized in Algorithm~\ref{alg:constraint-graph}.

\begin{algorithm}[htbp]

\caption{Constraint Graph Construction}
\label{alg:constraint-graph}
\begin{algorithmic}[1]
\Require Rectangle set $\mathcal R$ and its geometry before legalization.
\Ensure Horizontal and vertical constraint graphs $\mathcal G_{\mathrm H}$ and $\mathcal G_{\mathrm V}$.
\State Initialize $\mathcal N_{\alpha}$ and all edge subsets for $\alpha\in\{\mathrm H,\mathrm V\}$.
\State Add the source and sink edges of every $r\in\mathcal R$ to $\mathcal E_{\alpha}^{\mathrm I}$ for $\alpha\in\{\mathrm H,\mathrm V\}$.
\For{each unordered pair $\{u,v\}\subseteq\mathcal R$}
    \State Determine $\operatorname{dir}_{uv}$ using Eq.~\eqref{eq:constraint-direction-selection} and orient $(n_u,n_v)$ in direction $\operatorname{dir}_{uv}$.
    \State Add $(n_u,n_v)$ to $\mathcal E_{\operatorname{dir}_{uv}}^{\mathrm I}$.
\EndFor
\State $\mathcal E_{\alpha}^{\mathrm I}\gets\operatorname{TransitiveReduction}(\mathcal E_{\alpha}^{\mathrm I})$ for $\alpha\in\{\mathrm H,\mathrm V\}$.
\For{each soft module $v_i\in\mathcal V_s$}
    \For{each $(n_u,n_v)\in\mathcal E_{\mathrm H}^{\mathrm I}$ with $u,v\in\mathcal S_i$}
        \If{$\ell_{uv}^{\mathrm H}>0$}
            \State Move $(n_u,n_v)$ from $\mathcal E_{\mathrm H}^{\mathrm I}$ to $\mathcal E_{\mathrm H}^{=}$ and add the oriented pair to $\mathcal E_{\mathrm V}^{\mathrm O}$.
        \EndIf
    \EndFor
    \For{each $(n_u,n_v)\in\mathcal E_{\mathrm V}^{\mathrm I}$ with $u,v\in\mathcal S_i$}
        \If{$\ell_{uv}^{\mathrm V}>0$}
            \State Move $(n_u,n_v)$ from $\mathcal E_{\mathrm V}^{\mathrm I}$ to $\mathcal E_{\mathrm V}^{=}$ and add the oriented pair to $\mathcal E_{\mathrm H}^{\mathrm O}$.
        \EndIf
    \EndFor
\EndFor
\State \Return $\mathcal G_{\mathrm H}$ and $\mathcal G_{\mathrm V}$.
\end{algorithmic}
\end{algorithm}

Algorithm~\ref{alg:constraint-graph} constructs the two constraint graphs as follows. Lines 1--2 initialize $\mathcal G_{\mathrm H}$ and $\mathcal G_{\mathrm V}$ and connect every rectangle node to their boundary nodes. Lines 3--5 examine every rectangle pair, select its separation direction using Eq.~\eqref{eq:constraint-direction-selection}, and insert an inequality edge into $\mathcal E_{\mathrm H}^{\mathrm I}$ or $\mathcal E_{\mathrm V}^{\mathrm I}$. Line 6 removes redundant inequality edges by transitive reduction. Lines 7--13 then process the subrectangles of each soft module. Lines 8--10 convert every original left-right adjacency in $\mathcal G_{\mathrm H}$ into an equality edge and add its orthogonal overlap edge to $\mathcal G_{\mathrm V}$. Lines 11--13 symmetrically convert every original bottom-top adjacency in $\mathcal G_{\mathrm V}$ and add the corresponding overlap edge to $\mathcal G_{\mathrm H}$. Therefore, the inequality edges prevent overlap between modules, while the equality and orthogonal overlap edges retain the original adjacency and connectivity within each soft module. Line 14 returns the two graphs, whose typed edges are converted into affine constraints in the following convex formulation.
}

\subsubsection{Convex Geometric Legalization}
{
Given $\mathcal R$, its target areas, and the constraint graphs $\mathcal G_{\mathrm H}$ and $\mathcal G_{\mathrm V}$, we formulate geometric legalization as a convex optimization problem. For each $r\in\mathcal R$, let $(x_r,y_r,w_r,h_r)$ denote its center and dimensions. For a hard module, $w_r=\bar w_r$ and $h_r=\bar h_r$, while its center remains movable. For a preplaced module, $(x_r,y_r,w_r,h_r)=(\bar x_r,\bar y_r,\bar w_r,\bar h_r)$.

The typed graph edges are first imposed as affine topological constraints. For rectangles $u$ and $v$, define the center based gaps
\begin{equation}
\begin{aligned}
g_{uv}^{\mathrm H}&:=x_v-x_u-\frac{w_u+w_v}{2},\\
g_{uv}^{\mathrm V}&:=y_v-y_u-\frac{h_u+h_v}{2}.
\end{aligned}
\label{eq:constraint-graph-gap}
\end{equation}
The inter-module inequality edges and intra-module equality edges are enforced by
\begin{equation}
\begin{aligned}
g_{uv}^{\mathrm H}&\ge0,
&&\forall(n_u,n_v)\in\mathcal E_{\mathrm H}^{\mathrm I}, &
 g_{uv}^{\mathrm H}&=0,
&&\forall(n_u,n_v)\in\mathcal E_{\mathrm H}^{=},\\
g_{uv}^{\mathrm V}&\ge0,
&&\forall(n_u,n_v)\in\mathcal E_{\mathrm V}^{\mathrm I}, &
 g_{uv}^{\mathrm V}&=0,
&&\forall(n_u,n_v)\in\mathcal E_{\mathrm V}^{=}.
\end{aligned}
\label{eq:socp-separation-adjacency}
\end{equation}
For the orthogonal minimum-overlap relations, we impose
\begin{equation}
\begin{aligned}
&x_u+\frac{w_u}{2}-x_v+\frac{w_v}{2}
 \ge\epsilon_{\mathrm{ov}},
 \quad\forall(n_u,n_v)\in\mathcal E_{\mathrm H}^{\mathrm O},\\
&x_v+\frac{w_v}{2}-x_u+\frac{w_u}{2}
 \ge\epsilon_{\mathrm{ov}},
 \quad\forall(n_u,n_v)\in\mathcal E_{\mathrm H}^{\mathrm O},\\
&y_u+\frac{h_u}{2}-y_v+\frac{h_v}{2}
 \ge\epsilon_{\mathrm{ov}},
 \quad\forall(n_u,n_v)\in\mathcal E_{\mathrm V}^{\mathrm O},\\
&y_v+\frac{h_v}{2}-y_u+\frac{h_u}{2}
 \ge\epsilon_{\mathrm{ov}},
 \quad\forall(n_u,n_v)\in\mathcal E_{\mathrm V}^{\mathrm O}.
\end{aligned}
\label{eq:socp-minimum-overlap}
\end{equation}
where $\epsilon_{\mathrm{ov}}>0$ specifies the minimum shared-boundary length. Together, the equality and minimum-overlap constraints preserve the adjacency encoded by the decomposition.

For every soft subrectangle $S_i^j\in\mathcal S_i$, its width and height are variables constrained by
\begin{equation}
    w_i^jh_i^j\ge A_i^{j,\mathrm{tar}}.
    \label{eq:subrectangle-area}
\end{equation}
Let $\mathcal Q^{\mathrm{rot}}:=\{(a,b,z)\mid 2ab\ge z^2,\ a\ge0,\ b\ge0\}$ denote the rotated second-order cone. Constraint~\eqref{eq:subrectangle-area} is equivalently written as
\begin{equation}
\left(w_i^j,h_i^j,
\sqrt{2A_i^{j,\mathrm{tar}}}\right)\in\mathcal Q^{\mathrm{rot}}.
\label{eq:rotated-area-cone}
\end{equation}
To enforce the minimum inner width constraint~\eqref{eq:obj1:inner-width}, we additionally impose
\begin{equation}
    w_i^j\ge \omega^{\min},
    \qquad
    h_i^j\ge \omega^{\min},
    \label{eq:legalization-minimum-inner-width}
\end{equation}
for every soft subrectangle $S_i^j\in\mathcal S_i$, using the prescribed minimum inner width $\omega^{\min}$.

For each soft module $v_i$, let $(L_i,R_i,B_i,T_i)$ denote the bounding extents of its complete rectilinear shape. They are linked to every subrectangle by
\begin{equation}
\begin{aligned}
&L_i\le x_i^j-\frac{w_i^j}{2},
\quad R_i\ge x_i^j+\frac{w_i^j}{2},\\
&B_i\le y_i^j-\frac{h_i^j}{2},
\quad T_i\ge y_i^j+\frac{h_i^j}{2}.
\end{aligned}
\label{eq:module-extents}
\end{equation}
According to the aspect ratio convention $h_i/w_i$, the module level constraint is
\begin{equation}
\mathrm{AR}_i^l(R_i-L_i)
\le T_i-B_i
\le\mathrm{AR}_i^u(R_i-L_i).
\label{eq:socp-module-aspect-ratio}
\end{equation}

The source and sink terminals bound every rectangle according to
\begin{equation}
\begin{aligned}
&x_r-\frac{w_r}{2}\ge x_{s_{\mathrm H}},
\quad x_r+\frac{w_r}{2}\le x_{t_{\mathrm H}},\\
&y_r-\frac{h_r}{2}\ge y_{s_{\mathrm V}},
\quad y_r+\frac{h_r}{2}\le y_{t_{\mathrm V}}.
\end{aligned}
\label{eq:socp-envelope}
\end{equation}
In the fixed-outline mode, $(x_{s_{\mathrm H}},x_{t_{\mathrm H}},y_{s_{\mathrm V}},y_{t_{\mathrm V}})=(0,W,0,H)$. In the outline-free mode, these terminals define the variable outline width and height
\begin{equation}
W_{\mathrm{outline}}:=x_{t_{\mathrm H}}-x_{s_{\mathrm H}},
\qquad
H_{\mathrm{outline}}:=y_{t_{\mathrm V}}-y_{s_{\mathrm V}}.
\label{eq:outline-dimensions}
\end{equation}

For each rectangle $r\in\mathcal R$, its Manhattan displacement from the prelegalization position $(x_r^0,y_r^0)$ is represented by nonnegative epigraph variables satisfying
\begin{equation}
\begin{aligned}
&d_r^x\ge x_r-x_r^0,
\quad d_r^x\ge x_r^0-x_r,\\
&d_r^y\ge y_r-y_r^0,
\quad d_r^y\ge y_r^0-y_r.
\end{aligned}
\label{eq:socp-displacement}
\end{equation}

The convex formulation supports the two optimization modes defined in Section~\ref{sec:pre}. Under a fixed outline, it minimizes displacement with a weak compactness regularizer
\begin{equation}
\min\quad
\sum_{r\in\mathcal R}(d_r^x+d_r^y)
+\kappa\sum_{v_i\in\mathcal V_s}
\bigl[(R_i-L_i)+(T_i-B_i)\bigr].
\label{eq:fixed-outline-socp-objective}
\end{equation}
In the outline-free mode, chip area replaces displacement as the optimization objective
\begin{equation}
\min\quad
W_{\mathrm{outline}}+H_{\mathrm{outline}}
+\kappa\sum_{v_i\in\mathcal V_s}
\bigl[(R_i-L_i)+(T_i-B_i)\bigr],
\label{eq:outline-free-socp-objective}
\end{equation}
where $\kappa>0$ is a small weight that tightens the auxiliary module extents. The first objective preserves the global floorplanning solution, whereas the second minimizes chip area.
}

Finally, the optimized subrectangles belonging to the same soft module are merged to reconstruct its rectilinear polygon. Let $\mathcal P^*$ denote the resulting legalized polygon set. Together with the hard and preplaced modules, $\mathcal P^*$ constitutes the final legalized floorplan.

\begin{table*}[t]
  \centering
  \caption{Fixed-outline comparison on the GSRC and MCNC benchmark suites with 10\% whitespace (HPWL in $\mu$m and RT in seconds).}
  \resizebox{0.8\textwidth}{!}{%
    \begin{tabular}{ccc cc cc cc cc}
      \toprule
      \multicolumn{3}{c}{Benchmarks}
      & \multicolumn{2}{c}{SDP~\cite{2023SDP}}
      & \multicolumn{2}{c}{ICCAD'24~\cite{2024NTUFP}}
      & \multicolumn{2}{c}{DAC'26}
      & \multicolumn{2}{c}{Ours} \\
      \cmidrule(lr){1-3}\cmidrule(lr){4-5}\cmidrule(lr){6-7}\cmidrule(lr){8-9}\cmidrule(lr){10-11}
      Benchmark & \#Modules & \#Nets & HPWL & RT & HPWL & RT & HPWL & RT & HPWL & RT \\
      \midrule
      n10   & 10  & 118   & 36,277  & N/A & 32,772          & 0.52  & 33,036  & 1.84  & \textbf{32,591.62}  & 1.79 \\
      n30   & 30  & 349   & 106,013 & N/A & 99,406          & 2.85  & 96,885  & 2.77  & \textbf{95,756.00}  & 2.84 \\
      n50   & 50  & 485   & 136,547 & N/A & 120,526         & 3.52  & 121,149 & 3.56  & \textbf{118,916.28} & 3.65 \\
      n100  & 100 & 885   & 228,040 & N/A & 193,064         & 6.83  & 186,092 & 6.94  & \textbf{182,268.07} & 6.26 \\
      n200  & 200 & 1,585 & 407,091 & N/A & 355,313         & 42.92 & 338,766 & 15.40 & \textbf{330,791.62} & 10.40 \\
      n300  & 300 & 1,893 & N/A     & N/A & 479,704         & 62.34 & 465,695 & 24.76 & \textbf{442,964.71} & 13.70 \\
      ami33 & 33  & 122   & N/A     & N/A & 44,579          & 1.42  & 42,784  & 2.48  & \textbf{41,445.61}  & 2.58 \\
      ami49 & 49  & 396   & N/A     & N/A & 662,432         & 3.25  & 576,876 & 4.06  & \textbf{542,889.18} & 4.72 \\
      \midrule
      \multicolumn{3}{c}{Ratio}
      & 1.17 & N/A & 1.07 & 1.10 & 1.03 & 1.12 & 1.00 & 1.00 \\
      \bottomrule
    \end{tabular}%
  }
  \label{tab:fixed-outline-comparison}
\end{table*}

\section{Experimental Results}\label{sec:exp}
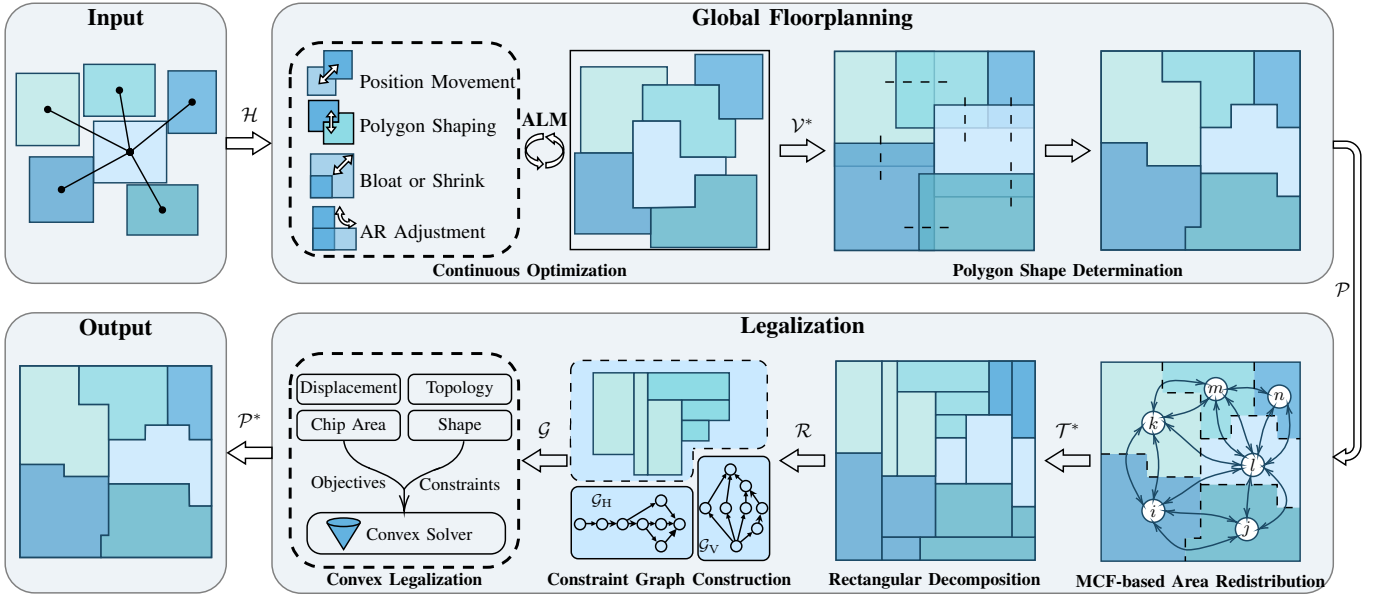
In this section, we present a comprehensive evaluation of the proposed floorplanning algorithm.
Our implementation uses C++ and several libraries, including \texttt{Libtorch}~\cite{paszke2019pytorch} for automatic differentiation and optimization, \texttt{GTL}~\cite{boost} for data structures and polygon operations, \texttt{Lemon}~\cite{dezsHo2011lemon} for network-flow optimization, and \texttt{MOSEK}~\cite{mosekfusion} for convex geometric legalization.
All experiments are executed on a machine equipped with an \texttt{AMD EPYC 7763} CPU @ 2.45 GHz and 128 GB RAM. All reported results are obtained on the CPU using eight threads.
The experimental settings used in our method are summarized in Table~\ref{tab:hyperparameters}. For the fixed-outline experiments, the outline dimensions are calculated as
\begin{equation}
  W = \sqrt{(1+\Gamma)A\xi}, \qquad
  H = \sqrt{(1+\Gamma)A/\xi},
  \label{eq:experimental-outline}
\end{equation}
where $A=\sum_{i=1}^{m}A_i^{\min}$ is the total minimum module area, $\xi$ is the outline aspect ratio, and $\Gamma$ is the whitespace ratio.

\begin{table}[htbp]
    \centering
    \caption{Experimental Settings}
    \resizebox{\hsize}{!}{
        \begin{tabular}{lll}
            \toprule
            Para.          & Description                            & Value                                                              \\
            \hline
            $\mathbf v^0$ & Initial BBox variables                 & $\{(\frac{W}{2},\frac{H}{2}, A_i^{\min},1)\mid v_i\in\mathcal V_s\}$ \\
            $\mathbf r^0$ & Initial shaping variables              & $\{0 \mid 1\le i < j \le m\}$                                      \\
            $\boldsymbol\lambda^0$    & Initial Lagrangian multiplier          & $\{0 \mid v_i\in\mathcal V_m\}$                                                                  \\
            $\mu_0$        & Initial penalty parameter              & 1.0                                                                \\
            $lr_0$         & Initial learning rate                  & $0.001W$                                                         \\
            $\tau$         & Penalty factor                         & 2.0                                                                \\
            $\theta$       & Scale factor                           & 0.95                                                                \\
            $\varepsilon_t$   & Target tolerance                       & 0.01                                                               \\
            N              & Maximum ALM iterations                & 100                                                                \\
            $N_{\mathrm{in}}$ & Maximum primal step iterations & 100                                                                \\
            Opt.           & Optimizer for primal step              & \texttt{torch::Optim::Adam}~\cite{paszke2019pytorch}                                        \\
            MCF Sol.       & Minimum-cost flow solver               & \texttt{lemon::NetworkSimplex}~\cite{dezsHo2011lemon} \\
            Convex Sol.    & Convex legalization solver             & \texttt{mosek::fusion}~\cite{mosekfusion} \\
            \bottomrule
        \end{tabular}
    }
    \label{tab:hyperparameters}
\end{table}

\subsection{Fixed-Outline Performance Comparison}
We first evaluate fixed-outline performance against SDP~\cite{2023SDP} and ICCAD'24~\cite{2024NTUFP} on the GSRC~\cite{GSRCbench} and MCNC~\cite{MCNCbench} benchmark suites, where all modules are modeled as rectilinear soft modules. We set the outline aspect ratio to $\xi=1$ and the whitespace ratio to $\Gamma=10\%$. ICCAD'24 and Ours set the rectangle ratio constraint of every module to $\mathrm{RR}=0.8$, while SDP does not support this constraint and is evaluated using its reported configuration.

For SDP, we report the available HPWL values from its paper and omit its runtime because of the high cost of the semidefinite programming solver. Since ICCAD'24 does not publish its hyperparameters, we execute its binary and follow its grid-search tuning strategy. Each benchmark is evaluated over 50 trials, and the best HPWL and its corresponding runtime are reported.

Table~\ref{tab:fixed-outline-comparison} summarizes the fixed-outline results. Because the benchmark scales differ substantially, the HPWL and RT ratios are calculated as geometric means of the per-instance ratios normalized to Ours over the commonly available instances. Since SDP reports HPWL only for \texttt{n10}--\texttt{n200} and does not report runtime, its HPWL ratio uses these five instances and its RT ratio is marked as N/A. The ICCAD'24 ratio uses all eight instances. Ours achieves the lowest HPWL on all eight benchmarks. The HPWL ratios are 1.17 and 1.07 for SDP and ICCAD'24, respectively, corresponding to geometric-mean HPWL reductions of 14.42\% and 6.49\%. The RT ratio of ICCAD'24 is 1.10, indicating that Ours also reduces geometric-mean runtime by 9.30\%.

\subsection{ICCAD 2023 contest Benchmark Comparison}
We further evaluate the proposed method on the ten fixed-outline instances of the ICCAD 2023 contest benchmark suite~\cite{cadcontest23}. The comparison includes the contest first-place result and ICCAD'24~\cite{2024NTUFP}. Following the legal-result comparison in ICCAD'24, the first-place HPWL values are measured after enforcing the minimum inner width constraint~\eqref{eq:obj1:inner-width}. The provided first-place binary uses a fixed 25-minute budget for every instance, rather than reporting a directly comparable algorithm runtime. Its RT entries are therefore marked as N/A.

\begin{table*}[t]
  \centering
  \caption{Comparison on the ICCAD 2023 contest benchmark suite (HPWL in $\mu$m and RT in seconds).}
  \resizebox{0.95\hsize}{!}{%
    \begin{tabular}{ccccc cc cc cc}
      \toprule
      \multicolumn{5}{c}{Benchmarks}
      & \multicolumn{2}{c}{1st Place~\cite{cadcontest23}}
      & \multicolumn{2}{c}{ICCAD'24~\cite{2024NTUFP}}
      & \multicolumn{2}{c}{Ours} \\
      \cmidrule(lr){1-5}\cmidrule(lr){6-7}\cmidrule(lr){8-9}\cmidrule(lr){10-11}
      Benchmark & \#Soft & \#Fixed & \#Nets & Outline
      & HPWL & RT & HPWL & RT & HPWL & RT \\
      \midrule
      case01 & 15 & 5  & 45  & 11,267$\times$10,450 & 144,074,741.0 & N/A & 149,021,269.0 & 1.60 & \textbf{138,545,842.0} & 1.85 \\
      case02 & 16 & 8  & 39  & 2,300$\times$2,300   & 18,366,208.0  & N/A & 19,170,372.0  & 1.44 & \textbf{17,318,640.0}  & 2.68 \\
      case03 & 28 & 14 & 108 & 2,500$\times$3,000   & 1,813,731.0   & N/A & 1,856,708.5   & 2.25 & \textbf{1,747,110.5}   & 2.54 \\
      case04 & 20 & 8  & 47  & 4,995$\times$4,407   & 60,376,550.0  & N/A & 57,433,825.0  & 1.52 & \textbf{52,907,575.0}  & 2.37 \\
      case05 & 16 & 8  & 33  & 4,620$\times$3,740   & 14,706,300.0  & N/A & 14,544,050.0  & 1.38 & \textbf{13,659,300.0}  & 1.73 \\
      case06 & 21 & 13 & 56  & 3,000$\times$2,700   & 36,642,600.0  & N/A & 34,354,550.0  & 1.53 & \textbf{33,408,700.0}  & 1.02 \\
      case07 & 16 & 11 & 39  & 12,200$\times$12,400 & 379,504,500.0 & N/A & 367,086,150.0 & 1.39 & \textbf{349,013,600.0} & 2.28 \\
      case08 & 37 & 4  & 72  & 21,500$\times$14,480 & 97,965,450.0  & N/A & 96,866,700.0  & 2.44 & \textbf{95,786,100.0}  & 1.87 \\
      case09 & 14 & 7  & 19  & 22,570$\times$18,200 & 90,645,250.0  & N/A & 97,915,250.0  & 1.32 & \textbf{76,708,750.0}  & 1.00 \\
      case10 & 7  & 5  & 11  & 13,510$\times$13,400 & 17,145,500.0  & N/A & \textbf{15,612,100.0} & 1.21 & 15,801,100.0 & 1.18 \\
      \midrule
      \multicolumn{5}{c}{Ratio}
      & 1.08 & N/A & 1.07 & 0.90 & 1.00 & 1.00 \\
      \bottomrule
    \end{tabular}%
  }
  \label{tab:iccad-contest-comparison}
\end{table*}

As shown in Table~\ref{tab:iccad-contest-comparison}, Ours achieves the lowest HPWL on nine of the ten instances, while ICCAD'24 gives the best result on \texttt{case10}. The geometric-mean HPWL ratios normalized to Ours are 1.08 and 1.07 for the contest first-place result and ICCAD'24, respectively. These values correspond to HPWL reductions of 7.58\% and 6.78\% by Ours. The RT ratio of ICCAD'24 is 0.90, while the contest first-place RT ratio is unavailable. Thus, the proposed method improves solution quality across most contest instances with an 11.39\% increase in geometric-mean runtime relative to ICCAD'24.

\subsection{Outline-Free Performance Comparison}
We next evaluate the outline-free mode against TOFU~\cite{2023tofu} and JigsawPlanner~\cite{2024jigsawplanner} on the GSRC and MCNC benchmark suites. These methods optimize wirelength while minimizing chip area. Following JigsawPlanner, the whitespace ratio (WS) is defined as the ratio of whitespace to chip area. In our method, the fixed-outline displacement objective is disabled, and chip area is minimized during convex geometric legalization.

\begin{table*}[t]
  \centering
  \caption{Outline-free comparison on the GSRC and MCNC benchmark suites.}
  \resizebox{0.75\textwidth}{!}{%
    \begin{tabular}{l ccc ccc ccc}
      \toprule
      \multirow{2}{*}{Benchmark}
      & \multicolumn{3}{c}{TOFU~\cite{2023tofu}}
      & \multicolumn{3}{c}{JigsawPlanner~\cite{2024jigsawplanner}}
      & \multicolumn{3}{c}{Ours} \\
      \cmidrule(lr){2-4}\cmidrule(lr){5-7}\cmidrule(lr){8-10}
      & HPWL & WS & RT(s) & HPWL & WS & RT(s) & HPWL & WS & RT(s) \\
      \midrule
      n10   & 35,210    & 7.30\% & \textbf{1.86} & 34,360  & $\le1\%$ & 10.22 & \textbf{32,325.21}  & \textbf{0.35\%} & 2.03 \\
      n30   & 109,190   & 3.33\% & \textbf{3.54} & 97,410  & $\le1\%$ & 11.34 & \textbf{94,342.66}  & \textbf{0.31\%} & 2.93 \\
      n50   & 147,650   & 5.22\% & \textbf{7.43} & 122,355 & $\le1\%$ & 11.50 & \textbf{117,939.58} & \textbf{0.34\%} & 5.78 \\
      n100  & 250,940   & 5.93\% & 24.32 & 185,600 & $\le1\%$ & 15.05 & \textbf{181,477.12} & \textbf{0.37\%} & \textbf{11.95} \\
      n200  & 484,510   & 4.77\% & 87.07 & 337,260 & $\le1\%$ & 20.73 & \textbf{329,239.83} & \textbf{0.08\%} & \textbf{14.61} \\
      n300  & 712,550   & 5.04\% & 197.25 & 451,680 & $\le1\%$ & 26.33 & \textbf{446,061.07} & \textbf{0.27\%} & \textbf{18.07} \\
      ami33 & 71,070    & 2.36\% & \textbf{2.87} & 48,990  & $\le1\%$ & 9.40  & \textbf{43,199.70}  & \textbf{0.36\%} & 3.00 \\
      ami49 & 1,144,620 & 1.25\% & \textbf{5.99} & 664,980 & $\le1\%$ & 18.79 & \textbf{571,959.51} & \textbf{0.07\%} & 5.80 \\
      \midrule
      Ratio & 1.42 & -- & 1.92 & 1.06 & -- & 2.38 & 1.00 & -- & 1.00 \\
      \bottomrule
    \end{tabular}%
  }
  \label{tab:outline-free-comparison}
\end{table*}

Table~\ref{tab:outline-free-comparison} compares HPWL, whitespace ratio, and runtime. JigsawPlanner first derives a fixed outline by setting the target whitespace to 0\% and then assigns discrete grids to modules within this outline. Since the discretization can introduce small module-area mismatches, its WS is conservatively reported as below 1\%. Ours achieves WS values between 0.07\% and 0.37\% and the lowest HPWL on every benchmark. The geometric-mean HPWL ratios of TOFU and JigsawPlanner normalized to Ours are 1.42 and 1.06, corresponding to HPWL reductions of 29.77\% and 5.65\%, respectively. The corresponding RT ratios are 1.92 and 2.38, indicating runtime reductions of 47.97\% and 57.96\%.

\input{include/Figure/outline_free_result_runtime.tex}

Figure~\ref{fig:outline-free-result-runtime} shows an outline-free floorplanning result and the average runtime distribution of our method. Global floorplanning and legalization account for 77.97\% and 22.03\% of the runtime on average, respectively.

\begin{table}[t]
  \centering
  \caption{Ablation study of global floorplanning. All results are geometric-mean ratios normalized to the full method.}
  \label{tab:global-ablation}
  \resizebox{\columnwidth}{!}{%
    \begin{tabular}{lcccc}
      \toprule
      \multirow{2}{*}{Configuration}
      & \multicolumn{2}{c}{ICCAD 2023 contest}
      & \multicolumn{2}{c}{GSRC and MCNC} \\
      \cmidrule(lr){2-3}\cmidrule(lr){4-5}
      & HPWL & RT & HPWL & RT \\
      \midrule
      Full                           & 1.00 & 1.00 & 1.00 & 1.00 \\
      Fixed Area                     & 1.02 & 0.96 & 1.03 & 0.91 \\
      Fixed Shaping                  & 1.07 & 1.03 & 1.02 & 1.09 \\
      RR-0.7                         & 0.96 & 1.03 & 0.96 & 1.07 \\
      RR-0.9                         & 1.05 & 1.10 & 0.99 & 1.09 \\
      QPM                            & 1.04 & 1.01 & 1.04 & 0.99 \\
      \bottomrule
    \end{tabular}%
  }
\end{table}

\subsection{Ablation Study}
We conduct an ablation study to evaluate the main components and geometric settings of global floorplanning. Table~\ref{tab:global-ablation} reports geometric-mean HPWL and runtime ratios normalized to the full method on the ICCAD 2023 contest and combined GSRC and MCNC benchmarks. The ablation configurations are summarized as follows.
\begin{itemize}
  \item \textbf{Fixed Area}: Fix the bounding-box area of every soft module at its allowed maximum instead of optimizing it.
  \item \textbf{Fixed Shaping}: Fix all pairwise shaping proportions at 0.5.
  \item \textbf{RR-0.7}: Set the minimum rectangle ratio $\mathrm{RR}^{\min}$ to 0.7.
  \item \textbf{RR-0.9}: Set the minimum rectangle ratio $\mathrm{RR}^{\min}$ to 0.9.
  \item \textbf{QPM}: Replace the augmented Lagrangian formulation with a quadratic penalty method while retaining the remaining settings.
\end{itemize}

Fixing the module areas increases HPWL by 1.60\% on the ICCAD 2023 contest benchmarks and 2.72\% on GSRC and MCNC, confirming the benefit of optimizing bounding-box areas during global floorplanning. Fixing the shaping proportions causes HPWL increases of 6.50\% and 1.86\%, respectively, and also increases runtime on both benchmark groups. This result shows that adaptively allocating pairwise overlaps is particularly important for instances with irregular outlines and preplaced modules. Replacing ALM with QPM degrades HPWL by 4.23\% on the ICCAD 2023 contest benchmarks and 3.60\% on GSRC and MCNC, while providing no consistent runtime reduction. This comparison demonstrates that the multiplier updates of ALM improve constraint handling beyond a fixed quadratic penalty.

The rectangle-ratio variants reveal the tradeoff between shape flexibility and solution quality. Relaxing $\mathrm{RR}^{\min}$ to 0.7 improves HPWL by 4.18\% and 3.66\% on the two benchmark groups, respectively, because greater rectilinear flexibility enlarges the feasible shaping space. Tightening $\mathrm{RR}^{\min}$ to 0.9 increases HPWL by 4.71\% on the ICCAD 2023 contest benchmarks, although it yields a 0.93\% reduction on GSRC and MCNC. Both variants therefore indicate that the rectangle-ratio constraint should be selected according to geometric requirements rather than runtime alone.

\section{Conclusion}\label{sec:con}
This paper presented RPolyPlan, a unified framework that incorporates rectilinear soft modules into both global floorplanning and legalization. A structured continuous representation and differentiable polygon shaping model enable module locations and rectilinear shapes to be jointly optimized. The augmented Lagrangian formulation effectively handles the coupled constraints, while the minimum-cost flow and convex legalization produce feasible solutions with limited displacement from the global floorplan. Experimental results on standard benchmarks show that RPolyPlan reduces HPWL over the evaluated state-of-the-art methods by 7\% and 6\% in fixed-outline and outline-free experiments, respectively, while minimizing chip area and maintaining practical runtime. These results demonstrate the effectiveness of exploiting rectilinear shape flexibility throughout the floorplanning flow.

\section*{Acknowledgments}
The authors would like to thank Prof. Iris Hui-Ru Jiang for sharing the binaries and providing technical details related to \cite{2024NTUFP}.
{ The authors disclose the use of generative AI tools for language editing and sentence refinement to enhance readability and clarity. All technical content, methodology, and results were produced and verified by the authors.}

\bibliographystyle{IEEEtran}
\bibliography{ref}

\end{document}